\documentclass[1p,10pt,onecolumn]{elsarticle}

\usepackage{listings} 
\usepackage{amssymb}

\usepackage{graphicx}
\usepackage[utf8]{inputenc}
\usepackage{fourier} 
\usepackage{array}
\usepackage{makecell}
\usepackage{caption}
\usepackage{tabularx}
\usepackage{subcaption}
\usepackage{amsmath}
\usepackage{hyperref}

\usepackage[
    type={CC},
    modifier={by-nc-nd},
    version={3.0},
]{doclicense}

\newsavebox{\twosubbox}

\begin{document}
\begin{frontmatter}

\title{Interactive Semi-automated Specification Mining for Debugging: An Experience Report}

\author[AB]{Mohammad Jafar Mashhadi}
\ead{mohammadjafar.mashha@ucalgary.ca}
\author[MB]{Taha R. ~Siddiqui}
\ead{trsiddiqui1989@gmail.com}
\author[AB]{Hadi Hemmati}
\ead{hadi.hemmati@ucalgary.ca}
\author[MP]{Howard Loewen}
\ead{hloewen@micropilot.com}

\address[AB]{Department of Electrical \& Computer Engineering, University of Calgary, AB, Canada}
\address[MB]{InfoMagnetics Technologies Corp, Winnipeg, MB, Canada}
\address[MP]{MicroPilot Inc., Winnipeg, MB, Canada}

\begin{abstract}
\textbf{Context:} Specification mining techniques are typically used to extract the specification of a software in the absence of (up-to-date) specification documents. This is useful for program comprehension, testing, and anomaly detection. However, specification mining can also potentially be used for debugging, where a faulty behavior is abstracted to give developers a context about the bug and help them locating it.  \textbf{Objective:} In this project, we investigate this idea in an industrial setting. We propose a very basic semi-automated specification mining approach for debugging and apply that on real reported issues from an AutoPilot software system from our industry partner, MicroPilot Inc. The objective is to assess the feasibility and usefulness of the approach in a real-world setting. \textbf{Method:} The approach is developed as a prototype tool, working on C code, which accept a set of relevant state fields and functions, per issue, and generates an extended finite state machine that represents the faulty behavior, abstracted with respect to the relevant context (the selected fields and functions). \textbf{Results:} We qualitatively evaluate the approach by a set of interviews (including observational studies) with the company's developers on their real-world reported bugs. The results show that a) our approach is feasible, b) it can be automated to some extent, and c) brings advantages over only using their code-level debugging tools. We also compared this approach with traditional fully automated state-merging algorithms and reported several issues when applying those techniques on a real-world debugging context. \textbf{Conclusion:} The main conclusion of this study is that the idea of an ``interactive'' specification mining rather than a fully automated mining tool is NOT impractical and indeed is useful for the debugging use case.

\end{abstract}


\end{frontmatter}

\section{Introduction}

The area of specification mining is not a new research topic and hence there are many studies that have focused on the problem of automatically extracting software specification \citep{liu2011mining}. The techniques, in general, apply reverse engineering to abstract specification of the system from artifacts such as the source code, execution traces, or logs files. Broadly speaking, the techniques can be categorized in either static or dynamic analysis. In static analysis, the artifact, e.g., the source code, is analyzed without an actual execution of the system, but executions are needed to infer the behavior of software in dynamic analysis. 

The most well-known application of specification mining is perhaps program comprehension and while that is true \citep{Cook:1998}, these techniques can also be used for other purposes, such as requirement engineering \citep{RESpecMining}, anomaly detection \citep{anomalyValdes}, automated test generation \citep{testGenFraser}, program monitoring \citep{christodorescu2007mining}, and automated repair \citep{repairSpec}. In this research, we have targeted the ``debugging'' use case of specification mining. We exploit the capabilities of execution traces to collect as much information during an execution, and present it in an abstracted organized way for developers to have a consolidated but detailed view of the program. 

The application domain of our study is safety critical systems. Specifically, our case study is on an Autopilot software designed for Unmanned Aerial Vehicles (UAVs). This study is part of a larger collaborative research project with MicroPilot Inc. \citep{micropilotProject}, a commercial UAV (both hardware and software components) company. The debugging use case of specification mining was motivated by the company and is in their current interest toward higher quality software.

The study is in the form of an experience report, where the goal is not necessarily to advance the state of the art in specification mining or debugging,  
but rather demonstrate the feasibility and usefulness of an academic research topic (specification mining) on a relatively untouched use case, i.e., debugging, in a real-world setting (industrial system). 
 
Most related studies in the specification mining domain propose fully automated approaches to infer an Extended Finite State Machine (EFSM). These approaches combine a control-specific behaviour mining (e.g., kTails \citep{ktails})  with a data-specific behaviour mining (e.g., using Daikon \citep{daikon}). In this study, we evaluate a semi-automated interactive approach for mining EFSMs, where we automatically abstract the control flow but ask the user to interactively guide the tool to abstract the data-specific behaviour. To be specific, we ask the user to define the abstraction level and perspective of the derived models.

The reason for taking this option is that automating data-specific behaviour mining (i.e., in the EFSM case, creating state invariants automatically), needs a relatively large and diverse set of execution instances (several traces that all cover a particular state) so that the automated approach can generalize the individual instances to a pattern, as a constraint. However, in the debugging mode, the users typically runs very few tests to root cause the error.  Therefore, a typical automated data-specific behaviour mining approach is not applicable here. Thus as a first step toward a light-weight EFSM generator for debugging, we follow a less disruptive approach (not asking the users to run any extra tests for their debugging tasks) and evaluate the potentials of a semi-automated approach, in an industrial setup. Note that we do not claim that a fully automated approach is out of question, but since the semi-automated approach is less costly (requires no more executions than the failing tests), it is wise that before proposing more complex approaches one tries the simpler solutions to evaluate their potentials as well as setting up baselines for future research. 

To achieve the above objectives, our semi-automated approach includes four main steps:
\begin{enumerate}
  \item The perspective of the derived specification is set by a list of bug-related fields and functions to be monitored. The list can be selected automatically or manually. 
  \item The user defines the abstraction-level, interactively, by choosing the constraints to be monitored on each selected field.
  \item The tool traces the selected fields and functions from the buggy code.
  \item Execution traces are abstracted in the form of EFSMs using the defined constraints and existing state-merging algorithms. 
\end{enumerate}

It is also important to mention that the benefit of the semi-automated approach is its flexibility for generating models in different abstraction levels, with minimal cost.
For instance, making a high-level state machine for initial understanding of the context and then zooming-in to the defective area, just by changing the set of fields and functions. 
However, the potential caveat is the manual steps involved for selecting the fields, functions, and constraints. 
We argue that unlike the common belief, this is not an infeasible approach, in practice. 

To evaluate the idea and our claims, we have applied the approach to a case study of a commercial safety critical system with over 500K line of embedded C code. 
We conducted three phases of interviews (including observational studies), where a total of 8 developers from the company experienced with our tool to help debugging multiple bugs from a pool of 17 real bugs from their issue tracking systems. 
All steps of the project, starting from the use case motivation to implementation and to the evaluation, were done, collaboratively, with the company experts to assure the relevance, usefulness, and feasibility. 

In addition to the interviews, we also compared 5 model inference algorithms such as kTails, gkTail, and EDSM, implemented in Model INference Tool (MINT) \cite{Walkinshaw2016,MintGithub}. The evaluated techniques had several limitations and challenges, when applying to our debugging context, which we have categorized and reported in the paper.

\vspace{3mm}
 
Overall, the results show that: 
\begin{itemize}
    \item Our specification mining tool creates valid and correct models, in the given abstraction levels and perspectives.
    \item The models were welcomed by the developers, as a nice add-on to their debugging practice.
    \item The idea of an ``interactive'' specification mining rather than a fully automated mining tool is NOT impractical and indeed is useful for the debugging use case.    
    \item The evaluated alternatives have several challenges in the context of large scale system debugging, which make them, in their current state, unusable, in practice.    
\end{itemize}

In the rest of this paper, we first explain a motivating example in Section \ref{motivation}. Then we present background and related work in Section \ref{BKG-RW}. In Section \ref{methodology}, we explain our approach in details. Next in Section \ref{empiricalstudy}, we explain our empirical study. In Section \ref{futwork} we introduce some future paths that this study can be extended on then we conclude the paper in Section \ref{concl}. Finally, Section \ref{appendix} (Appendix), provides more details about the interview questions and answers.

\section{Motivation} \label{motivation}
There are many studies in academia that focus on algorithmic or technical aspects of specification mining. 
Two main techniques have been in the core of most of these approaches. 
These are kTails approach \citep{ktails} for state abstraction into Finite State Machines and invariant generation for data-related behaviour \citep{daikon}. 
We will explain them in more details in Section \ref{BKG-RW}, but in a nutshell the kTails-based approaches create the new states after (and/or before) any method/function call and then try to merge similar states. 
In invariant generation, however, a new state is defined based on the invariants (e.g., if a system variable X is negative then we are in State 1 otherwise in State 2). 
There are also techniques that combine both approaches to create Extended Finite State Machines (such as gkTail \citep{gk-tail}). 
Most of these work try hard to fully automatically come up with a precise general-purpose model of the system.

Our main argument for an interactive approach for specification mining is that in a context such as debugging, 
debuggers can afford some interactions with a specification miner, 
if that gives them more control and accuracy over what to monitor. 
In other words, we want to improve the current practice of debugging where logging, assertions, profiling, and breakpoints are mainly used. 
In this context, the full automation is less important, if the output models are not exactly depicting the problem. 
This can happen, for instance, with an invariant generator that simply hides the buggy behaviour into a larger state. 
Or with a kTails approach that floods the user with unwanted states after each method call.

To explain these limitations, let's first take an example of a generic autopilot software and demonstrate the system behaviour as depicted by a typical kTail-based specification mining approach and compare it with our proposed approach. 

As shown in the code snippet below, assume that a sequence of functions \textit{\{accelerate,takeoff\}} is called in a loop between the states of \texttt{Onground} and \texttt{Takenoff}. The \texttt{Onground} state is defined as \textit{altitude \textless 0} and \texttt{Takenoff} as \textit{altitude$\geq$0}. There might be many function calls that start from \texttt{Onground} but does not change it to \texttt{Takenoff}. In the above example, the \textit{takeoff} function, for instance, is called during the \texttt{Onground} state without any effect to the state, until the condition \textit{speed$\geq$ takeOffSpeed} is true.


\begin{lstlisting}[
		   basicstyle=\small,
		   language=C++,
                   directivestyle={\color{black}}
                   emph={int,char,double,float,unsigned},
                   emphstyle={\color{blue}},
                   label=code, 
                   caption=Example Code Snippet]
void main(){
    while(altitude<desiredAltitude){
        accelerate();
        takeOff();}
    while(!flight_success)
        fly();
    while(altitude>0){
        decelerate();
        land();}
    park();
}
void accelerate(){
    speed+=SPD_INCREMENT_CONST;
}
void takeOff(){
    if(speed>takeoffSpeed)
        while(altitude<desiredAltitude){
            altitude+=ALT_INCREMENT_CONST;
            if(altitude>safeAltForGearRetract)
                retractLandingGear();}
}
void decelerate(){
    speed-=SPD_INCREMENT_CONST;
}
void land(){
    if(speed <= landingSpeed)
        while(altitude>0){
            altitude-=ALT_INCREMENT_CONST;
            if(altitude<=safeAltForGearRetract)
                deployLandingGear();}
}
\end{lstlisting}

\begin{figure}
\centering
  \includegraphics[width=0.9\textwidth]{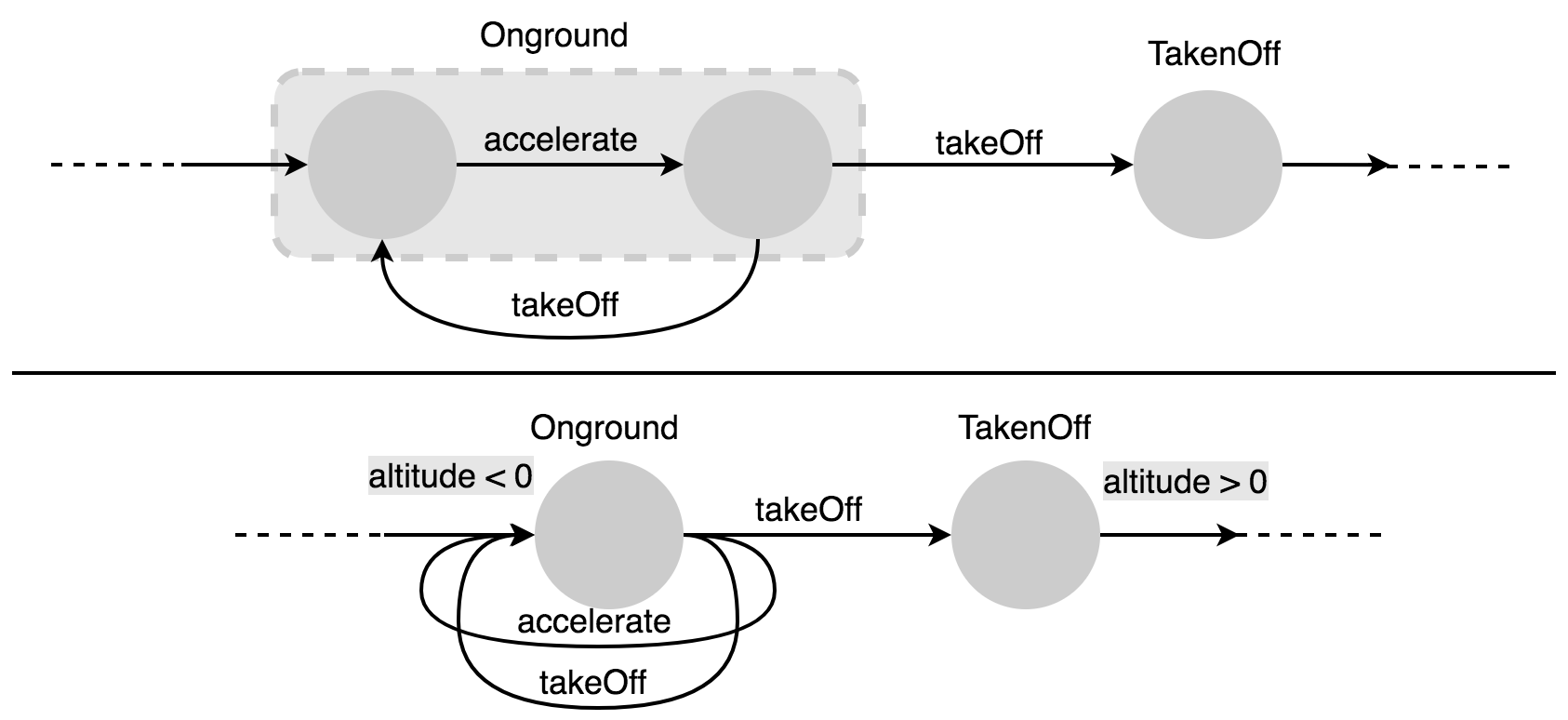}
  \caption{Sample extracted models using existing automated approaches (above) vs. our interactive approach (below).}
  \label{figure:comparison}
\end{figure}

Following a traditional kTails-based state merging strategy, the same function calls will be merged correctly into one transition but there will be two separate states for the \texttt{Onground} state, one after each call of \texttt{takeoff} and one after each call of \texttt{accelerate}. The partial state machine in Fig \ref{figure:comparison} (the upper one) shows this merging scenario using traditional kTail specification mining (where new function calls will result in new states). Assuming that the debugger is only interested in the two states of \texttt{Onground} and \texttt{Takenoff}, the automatically generated models are too detailed for this task.

Using automatically generated invariants helps in pruning such states but a tool like Daikon for invariant generation needs many execution traces to properly mine the patterns \citep{Walkinshaw2016}. This is not convenient for debugging where you want the limited QA budget to be spent on the actual testing and bug fixing rather than many system executions just to create an abstract model of a buggy behaviour. In addition, there is no guarantee that the output invariants are those constraints that a debugger is interested in.  The best level of abstraction for debugging is completely dependent on the bug's nature.  

Therefore, we propose to bring back the human in the loop. Following our interactive approach, we do not necessarily create states after/before each function call and rather make them just when the user defined constraints changes. So in that sense our approach is an Extended FSM (EFSM) since it uses data flow, as well. Though we do not generate explicit guard constraints at the moment.
Fig \ref{figure:comparison} shows the results of our interactive merging strategy, which is explained in detail in the methodology section (Section \ref{methodology}). 

\begin{figure}
  \includegraphics[width=0.9\textwidth]{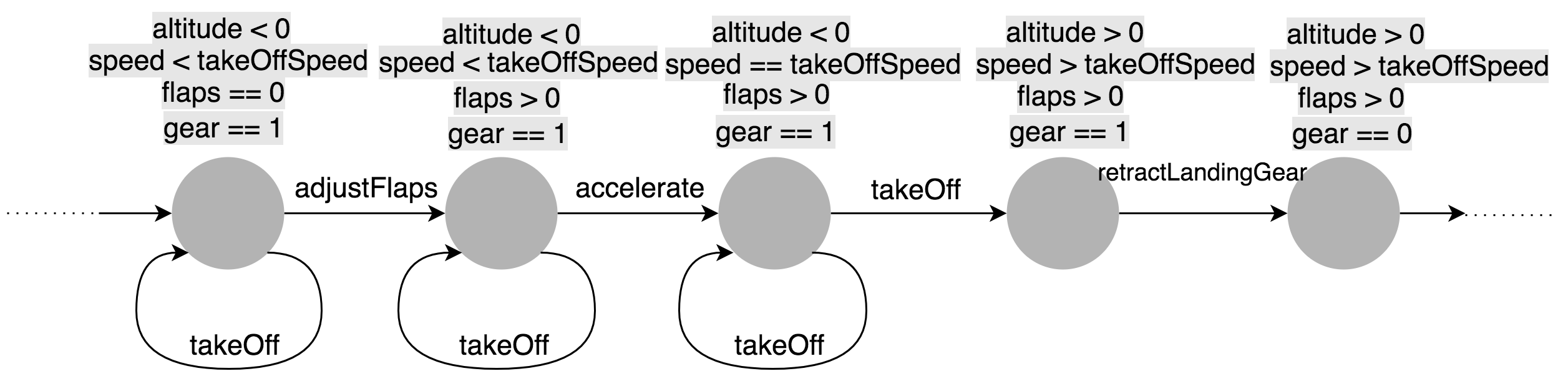}
  \caption{Focused state machine on Takeoff.}
  \label{figure:takeoff}
\end{figure}

\begin{figure}
\centering
  \includegraphics[width=0.9\textwidth]{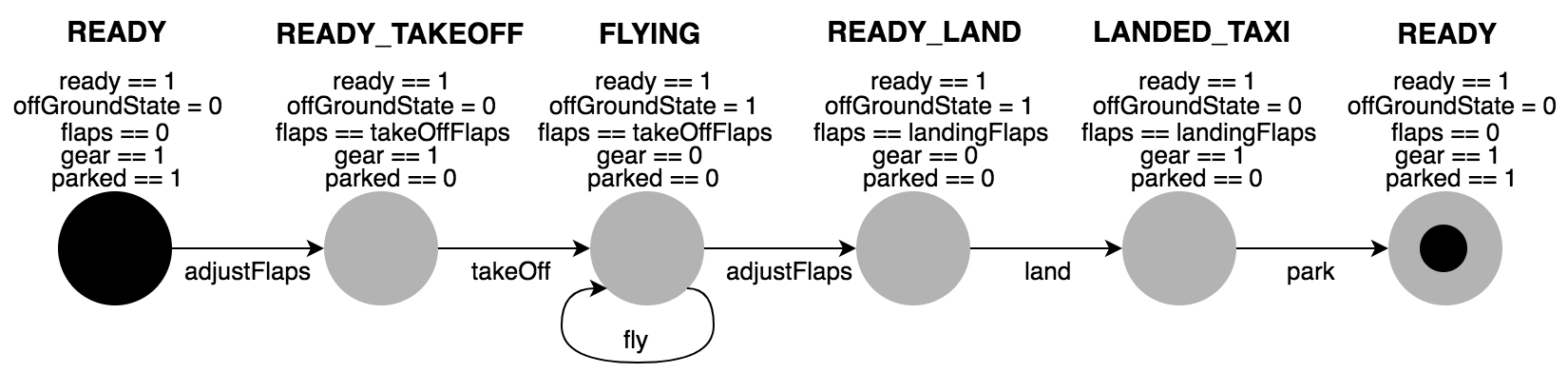}
  \caption{High-level state machine of a flight.}
  \label{figure:other}
\end{figure}

The main idea is to have the ability to define which states the user is interested in monitoring. This is helpful when one tries to locate a fault for debugging and knows the context from the bug report. This means that we need a partial model which only focuses on the state changes with respect to the relevant variables. To provide that, the interactive feature of our approach allows the user to create different models from the same set of execution traces, based on their needs. For instance, the state machine shown in Fig \ref{figure:takeoff} is generated using a set of four constraints and the generated state machine focuses only on the ``Take Off'' aspect of the flight. However, Fig \ref{figure:other} abstracts the same execution traces, in a higher level.

Finally, unlike the invariant-based approaches which need many traces to infer the patterns,  we can correctly and accurately abstract behaviour even with only a single execution scenario (corresponding to one reported bug). The user can also use multiple executions of the program and refine their models iteratively by adding and removing execution traces in the input as well as making adjustments in other parameters such as constraints and invariants. In addition, the level of granularity of the models is up to the user too. They can modify it by changing the functions they select for monitoring.

Note that we do not claim that a tailored version of any of the existing fully automated tools can not be used in this context. However, we opt to use an interactive approach, since we believe that the human expert is the best to decide about the perspective and level of abstraction, in the context of debugging. In the evaluation section, we will show that the extra cost of such decision (interaction overhead) is not a significant issue, specially in the context of debugging where the users are experts in the system and the bug reports are available.  

So to summarize, the motivation behind this work is to give up on the automation level of most existing work, to some extent, in favor of a special-purpose specification mining tool that is cheap (can operate even on only one bug's traces), accurate (constraints are provided not inferred), focused (partial models are generated based on the reported bug's perspective), and interactive (can give many views of the same behaviour from different perspectives and abstraction levels). 

In the following sections, we explain the details of this approach and through a real-world case study we evaluate its correctness, feasibility, cost, and usefulness. 

\section{Background and Related works} \label{BKG-RW}
In the past, several research projects have worked on the problem of specification mining \citep{softwareproductlines,modisco,legacysystem,walkinshawgrammerinference}.
Dynamic specification mining approaches, which are the context of this paper, work by executing the code and mining execution traces or logs. 

Execution traces typically consist of sequences of method calls, and other related information. 
These sequences can be generated by instrumenting the program and running the system with different inputs (different scenarios), the more the better, to cover the overall behavior of the system and hence producing correct and valid specification of the system. 
Higher coverage by the test inputs results in more accurate and complete specification models.
There are several studies that try to automatically create test inputs to properly capture the behavioral space \citep{dallmeier2010generating,pradel2012leveraging}.

Specification mining not only helps modelling the behavior of a software system, 
but is also extremely useful for a wide range of software engineering tasks, 
such as requirement generation \citep{ackermann2010automatic}, malware detection \citep{christodorescu2007mining}, 
anomaly detection \citep{cyber-attack}, and test case generation \citep{regressiontest}. 

In the rest of this section, we give a brief introduction to dynamic specification mining and traditional debugging techniques.
Other categories of related work that are slightly more remote and we do not explore in detail are 
extracting models for program comprehension \citep{debuggingsurvey}, 
anomaly detection based on dynamic analysis \citep{malik2013automatic}, and 
the static specification mining \citep{shoham2008static}.

\subsection{Dynamic Specification Mining Techniques}

Most dynamic specification mining techniques infer Finite State Automata or Finite State Machines (FSM) from execution traces.
These techniques begin by building a Prefix Tree Acceptor (PTA), 
a tree-shaped diagram of interconnected states showing the flow of a program in terms of method calls. 
PTA is generated from concrete execution traces, and contains the exact paths that are acceptable by the system. 
Since this step is a mere translation of traces to a tree, it contains a huge number of states and has recurring behavior all over (no abstraction). 

Thus abstraction techniques have been used to merge similar states in the mined FSMs. Dallmeier et. al. introduced a prototype called ADABU \citep{ADABU} for JAVA programs that mines models after classifying all the methods into two categories: 
Mutator (that changes the state of an observable object), and Inspector (that reveals properties of that object). 
Their mining approach dynamically captures the effect of mutator methods of the object state by calling all possible inspectors before and after calling each mutator. 
Therefore, the abstractions are based on the return values of inspectors. 
Our approach is in line with ADABU \citep{ADABU}, where for any function to be included as a transition, 
it has to be a \textit{mutator} and all the \textit{inspector}s are associated with states. 

Bierman et. al.'s kTails\citep{ktails} algorithm iteratively merges the pair of states in a state machine, if they are k-equivalent, until no such pair remains. 
K-equivalence here denotes that both states have the same sequence of leaving transitions of length k. 
The state machines generated by the above technique depict the program behavior, and apply merging techniques on a sequence of function calls. 

However, data constraints are important aspects of program behavior and if not included, the inferred model is deemed incomplete. 
In this regards, Extended Finite State Machines(EFSM) are helpful since they annotate the edges in FSM with data constraints valid for that interaction. 
gkTail \citep{gk-tail} is an example of abstraction algorithm that creates EFSMs, 
in which Lorenzoli et. al. adapted the kTails algorithm with information of data constraints observed during program interactions. 
gkTail merges similar states based on state invariants generated by Daikon \citep{daikon}. 

Daikon is a tool for the dynamic detection of likely invariants. It infers invariants by observing the variable values computed over a certain number of program executions. 
They can either be single value invariants ($a == 1$), or a relation between different properties ($x = y + z - 1$) that are helpful in defining behavior of the program with respect to its properties. 

Another usage of Daikon can be seen in \citep{lo2012scenario} in which Lo \& Maoz generated invariant augmented Live Sequence Charts.  

Lo et. al. \citep{lo2012learning} also empirically compared mining techniques that infer FSMs vs. those that mine EFSMs and reported that EFSM generators like gkTail are not necessarily more effective than FSM generators, specially gkTail was extremely slow.
This is inline with our motivation to propose an alternative that builds up on constraints (i.e., EFSM) but is very fast and thus more appropriate for the debugging context.

There are also techniques that are designed based on the same basic ideas (e.g., kTails), but propose more advanced algorithms. 
For example, Le et. al., \citep{le2015synergizing} propose a hybrid method to synergize existing specification miners to achieve higher accuracies. 
Krka et. al., \citep{krka2010using} propose another hybrid technique that uses execution traces and dynamically inferred invariants to overcome the over-generalization problem of kTail-based approaches.
Walkinshaw et. al. \citep{Walkinshaw2016} propose an alternative to invariant detection where a ``data classifier inference'' learns the class/output of variables from previous observations. 

User guided abstraction is an already well known idea. Previously Lo and Maoz has used user input during the abstraction to control the level of abstraction, filter the inputs, and selecting a subset of outputs for further investigations. \cite{Lo2008, Lo2009} Dallmeier et. al.'s ADABU \cite{ADABU} asked users for input to generate guards for its inferred models instead of automatically inferring them using a framework such as Daikon. Damas et. al.'s work on mining behavioural models also interacts with the user during the model generation to refine its output \cite{RESpecMining}. Walkinshaw et. al. called this method, asking user for input during the abstraction, active inference techniques \cite{walkinshaw2008inferring}. They ask users whether certain paths in the inferred model are valid paths in the actual software and they use the answer to treat the path as a positive example or a counter example of the software behaviour which is similar to Dupont et. al.'s QSM (Query-driven state merging) \cite{dupont2008qsm}. They have previously explored this idea in their 2007 paper ``Reverse engineering state machines by interactive grammar inference'' as well \cite{walkinshaw2007reverse}.

There are also several techniques that mine models other than state machines. 
For example, Lo et. al., \citep{lo2007mining} proposed a technique to mine Live Sequence Charts from execution traces. 
Lemieux et. al, \citep{lemieux2015general}, on the other hand, focus on the temporal properties and infer a linear temporal logic (LTL).

Finally, Pradel et. al. \citep{pradel2010framework},  provide a framework to compare different specification mining approaches. 
This is useful when there is no common ground to evaluate the effectiveness of the miners. 
However, in our case, we have the domain expert's opinion as the ground truth for both validation of the models and for evaluating their usefulness.

\subsection{Debugging}
Traditional debugging techniques include logging, assertion, profiling, and breakpoints.
Logging is performed by printing the program state or general messages through code instrumentation. 
Assertions are constraints, that are added to desired locations in a program, that when become false break the program. 
Profiling is the run-time analysis of performance of the program by monitoring its memory, CPU, etc. usage, 
which can help in detecting bugs such as memory leaks and unexpected calls to functions. \citep{surveyfaultlocalization}. 
However, in practice, breakpoints are the most common method for debugging, 
where the code is paused at the location of a breakpoint and the developer can inspect the variables and other context information at that instant of the execution. 
Effectiveness of breakpoints largely depends on the knowledge of the developer to put these breakpoints at relevant locations.

In this study, we take the combination of above traditional approaches (see Section \ref{empiricalstudy} for details) as the ``state of practice'' and compare our tool with. 
However, there are also more advanced debugging techniques in the academic literature that make use of execution traces to get insight from the execution of program, 
which can also be used for monitoring the system in run-time. 
Several development frameworks have logging capabilities which can be used by developers to log statements in certain events (e.g., warning and errors). 
However, the events are limited and require a lot of developer's effort to inspect the code, looking at a log statement. 

Spinellis \citep{debugbook} considers execution traces better than application-level logging for several reasons, 
especially for the frameworks that lack default logging capabilities, including the ability to debug the faulty situations in production environment after a full execution of the program.

Maoz \citep{moazmodelruntime} proposed the idea of using the abstracted traces as run-time models when a program is run against a defined high-level model, 
collecting relevant low-level data during the execution of the program. 
Maoz also used the idea of selective tracing, hence limiting the scope of traces to the scope of the higher-level models. 
The presented technique in this paper also uses the same basic idea of limiting the scope of execution traces to improve the efficiency of the tool 
and also help developers to focus on the selected context.

Another related category of studies is model-based debugging, 
where the  main idea is  automatically generating models from real execution of a program, 
and identify the location of the bugs by comparing its expected (given as models) and actual behavior (extracted) \citep{hybriddebugging}. 
Mayer et. al. \citep{stateoftheart} also used the same idea and applied artificial intelligence techniques on run-time models to automatically report suspected faults or assumptions in the case of deviating behavior. 
Other applications of this idea in the same category are also applied in \citep{hardwaredesign,javamodeling}. 
Shang et. al.\citep{shang2013assisting} used the same idea for debugging big data applications, 
but extracted models from Hadoop framework's log files rather than execution traces.

It is worth mentioning that there have been some research conducted in automatically finding the differences between state machines. For example, in the paper by Goldstein et. al. \cite{goldstein2017experience} they automatically compared two state machines, one representing the normal behaviour and the other representing the buggy behaviour of the program to find the differences in behaviour. The second state machine is not limited to being a buggy behaviour, it can be the model of a system after a software update and in this case, the difference can be used to find potential abnormal behaviour after the update, or a regression bug. Another example is Amar et. al.'s 2KDiff and its generalized version, nKDiff algorithms that find the differences between two program logs (translatable to the execution traces in our case) by making FSMs out of the logs and then comparing the state machines.

Some other techniques than can be useful for debugging using state models are those that treat state machines as directed graphs and automatically find different sub-graphs, assuming we have a base correct model before the bug. For instance, Cheng et. al. use a discriminating graph mining algorithm \cite{yan2008mining} to find top $k$ most discriminative sub-graphs that makes the faulty graph different from the ``normal'' ones \cite{cheng2009identifying}. Sun and Khoo also introduced a discriminative sub-graph finding algorithms that outperforms the previous work \cite{sun2013mining}. This sub-graph can be used as a \textit{bug-signature} \cite{hsu2008rapid}. Zuo et. al. also extracted another form of bug-signatures using hierarchical instrumentation \cite{zuo2014efficient}. In practice, these techniques can be useful for regression testing, but not all type of debugging.

\section{Approach} \label{methodology}
The approach presented in this paper proposes an interactive semi-automated technique to extract specification of a software system in the right level of details and for a given context. The main idea of the approach is to provide a light-weight model inference that keeps the human in the loop. Our approach includes nine steps divided into two phases: 

\begin{itemize}
    \item Execution trace extraction: 
        An interactive step to get the user input, in terms of what to monitor and in which level of detail.
    \item State Merging:
        Abstraction and merging of multiple executions in a concise state machine.
        
\end{itemize}

\begin{figure}
\centering
  \includegraphics[width=1 \textwidth]{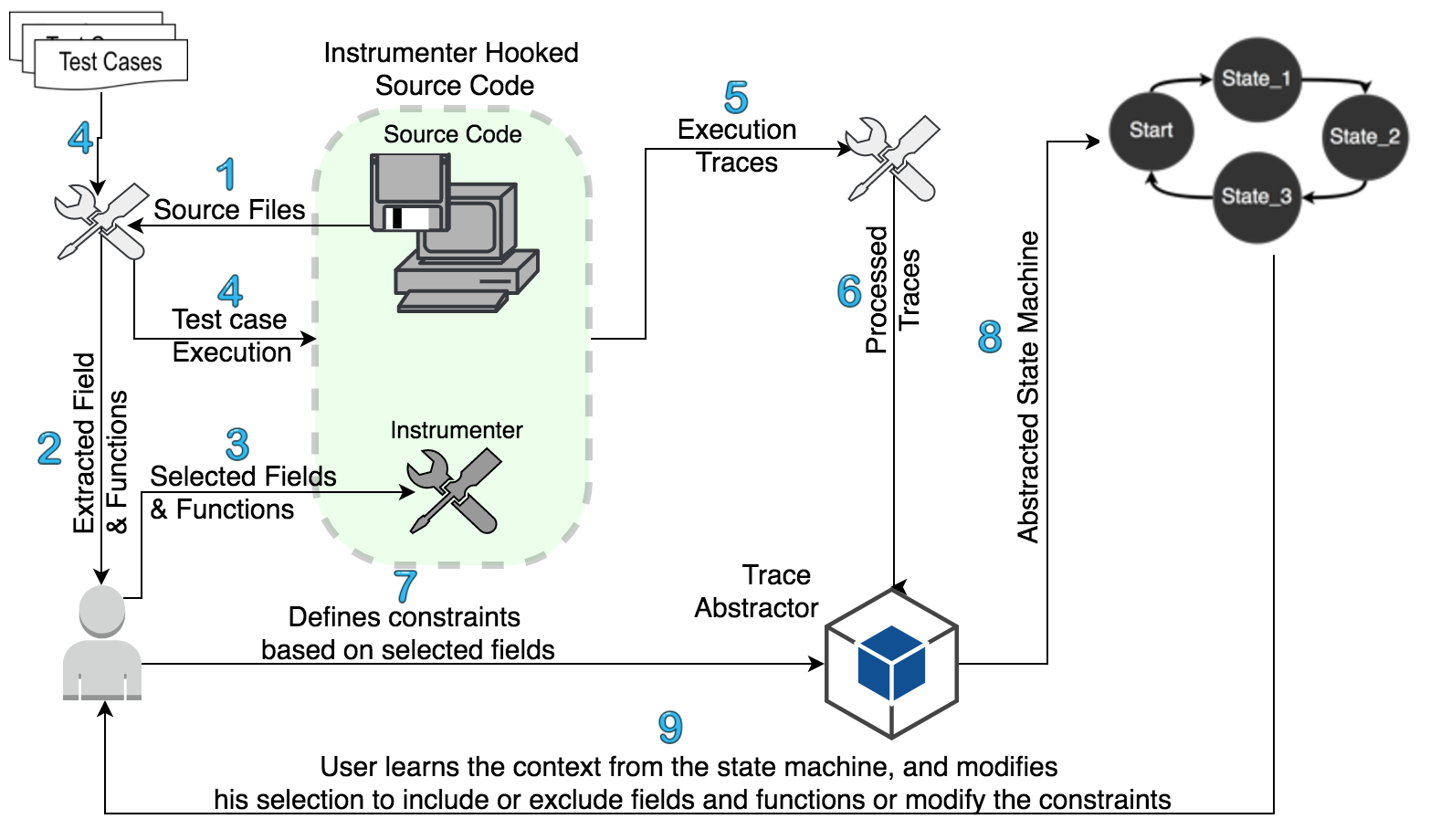}
  \caption{Overview of the proposed interactive specification mining approach}
  \label{figure:functionality}
\end{figure}

An overview of the proposed approach is shown in Fig \ref{figure:functionality}, where the details of the nine steps are explained in the subsections below.

\subsection{Execution Trace Extraction} \label{executionTraces}
The process starts with a static analysis tool that reads the raw source files, in \textbf{step 1}, from the subject source code and extracts a list of global fields and functions. The tool is based on \textit{Exuberant CTAGS}\citep{ctags}. For a given source file, ctags generates a list of tags for all global variables and functions in the code. The generated format is consistent to be read by any scripting tool. The tool generates tag files for each source file of the code-base and compiles a list of state fields by first analyzing the fields and structs, and then recursively analyzing the contained properties from tag files, if they are structs. 

In \textbf{step 2}, the extracted lists are displayed in an interactive GUI with several filters. The user selects the context of the state machine by selecting the related fields and functions with respect to his/her requirement (e.g., debugging a given bug). There is also an option of starting by loading a predefined list, which corresponds to a common aspect (e.g., takeOff) that have been used and saved, in the past.

In \textbf{step 3}, references to the selected fields and functions are passed to an instrumenter program, which is then compiled and linked with the subject code automatically. The purpose of this step is logging name of the functions, together with the values of global variables, that are called during execution of the code. Note that we only log global variables since we have been told by the company expert that in our case study this would be enough. However, logging local variables can also be easily added to the tool, following the same approach. 

We used the built-in compiler option in visual studio to attach hook functions to the subject code by enabling the development switch {\verb /Gh } and {\verb /GH }. This hooks the methods of {\verb _penter } and {\verb _pexit } to the entry and exit point of all functions without adding any additional line of code to the original code-base. By hooking the entry and exits of a function, we extract the context of the function, called during the execution; as the function name, parameter names and values, and the stack trace. 

We also trace the values of the selected global variables before entering and after exiting a function to notice the exact effect of the method call on the system. The code was implemented as a C++ static library that can be attached with any C/C++ application/dynamic library in visual studio to get execution traces. The library is compiled in released mode to work with the release version of the code which makes the execution faster, hence making the process efficient.

In \textbf{step 4}, our tool executes the code-base with a list of given test cases or scenarios. These are the test cases that fail or scenarios that are given in a bug report that reproduce a bug.
Once the code-base is executed the traces are gathered in \textbf{step 5}.

The raw log files of selected traces collected in \textbf{step 5} can be quite large in real world programs, due to many function calls that do not affect the selected fields. To minimize the traces, we have two options: a) to trace only those system interactions that result in change of state (fields), or b) to collect all details first and then process them to keep only the relevant selected functions (those that affect the selected fields) from each trace. 

The former option is more expensive due to the frequent comparison of state fields on all functions calls. Hence, in \textbf{step 6}, we take the later approach. Note that we do not discard the rest of the trace log file. Keeping all details will let us generate state machines of the same executions, but with different context (e.g., state fields to monitor) very fast. This is quite useful for debugging when the user may need to modify the selected fields several times to find the best state machine representing the exact context with all relevant details.  

The change impact analysis of function calls on state fields is done by tracing the entry and exit points of the functions, where we also get the information of nested calls. By comparing the ``before-enter'' and ``after-exit'' values of the fields, we can determine which function changed which fields. For the functions that contain nested calls, we can get the information by maintaining a stack of ``before-enter'' and ``after-exit'' values and compare the respective values. 

\subsection{State Merging} \label{intMerging}
Our proposed approach for state abstraction and merging is similar to the idea of ADABU, by Dallmeier et. al. \cite{ADABU}, where we let the users define constraints on the selected fields (\textbf{step 7} of Fig \ref{figure:functionality}).


The constraints are defined using a template selected from a predefined list of constraint templates. The list is a small subset of existing templates in the literature that are chosen with the industry partner's consultation. Note that the templates, which we explain them later, are very generic but they can be extended for other application domains, if necessary. The mining algorithms are not dependent on these specific templates. The approach is designed and implemented in a modular way so the templates can be modified for new applications without any changes being required to other parts. 

In practice, the user selects one or many state field(s) and their constraint template(s), individually.   
The concrete traces then will be abstracted to only show states that can be uniquely identified using the combination of these constraints. 
The current set of templates are described in Table \ref{constraintsTable}. 
Each template is explained with an example based on the example code discussed in Section \ref{motivation}.

\begin{table}[]
\caption{Categories of Merging Constraints}
\label{constraintsTable}
  \begin{center}
    \begin{tabular}{ | c | c | c | c |}
      \hline
       
\thead{Category} & \thead{Example}\\\hline

\makecell{ValueChange\\ of X}  &  \makecell{
      \centering
      \includegraphics[width=0.5\linewidth]{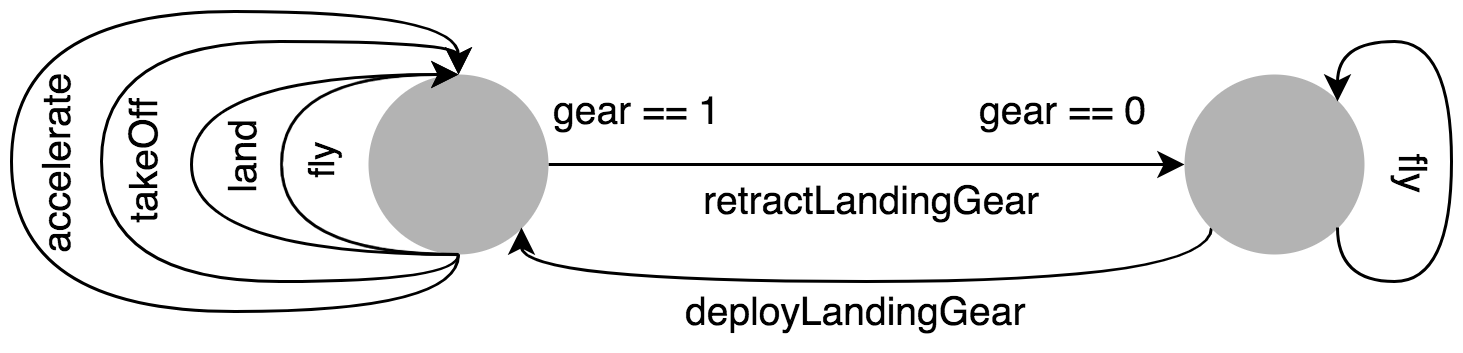} 
      } \\\hline

\makecell{X compared\\With Y or \\constant}  & \makecell{
      \centering
      \includegraphics[width=0.6\linewidth]{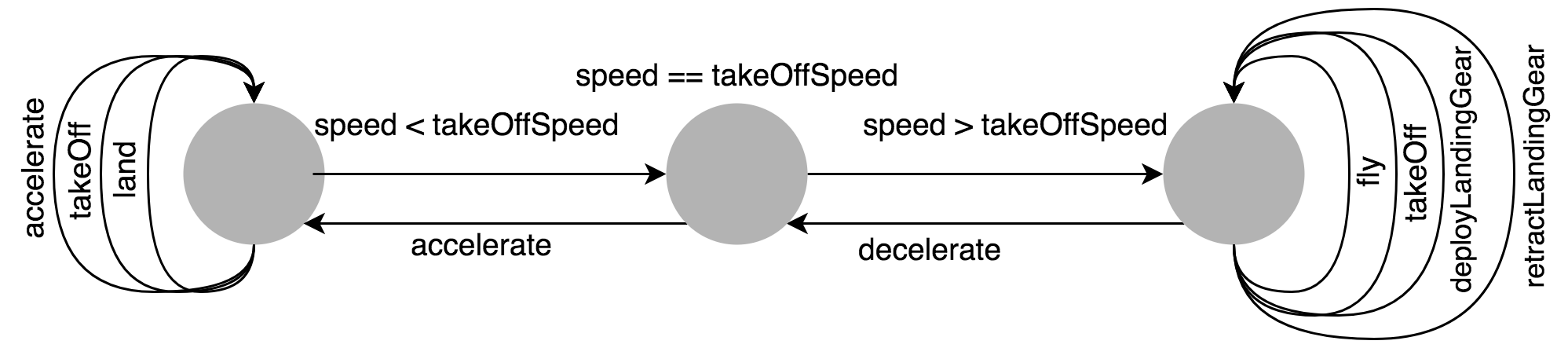}
      } \\\hline

\makecell{X compared\\With a Range\\(Y,Z) or\\(const,const)} & 
    \makecell{
      \centering
      \includegraphics[width=.8\linewidth]{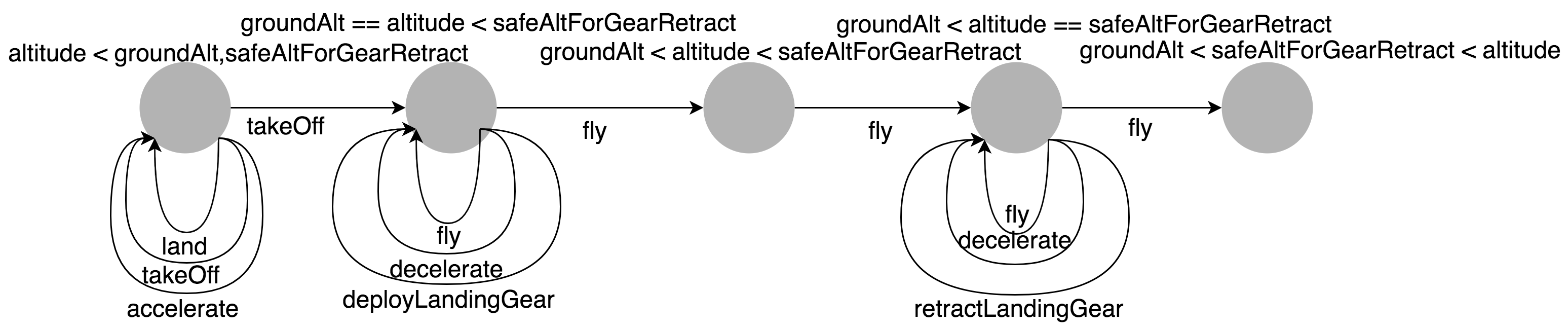}} \\\hline
       
    \end{tabular}
  \end{center}
\end{table}

\textbf{Template 1 (\texttt{ValueChange})}: This constraint accepts one state field (X). Based on this constraint, a new state is generated when the system detects a change in the value of X. This is an important category as most of the state dependent systems (specially in the embedded software domain) keep track of some internal states using several fields, where each value represents a unique state of the system. The corresponding example in Table \ref{constraintsTable} shows the states generated due to change in value of field \textit{gear} (\textit{gear==0} and \textit{gear==1}). Clearly, this template must not be used with continuous values. 

\textbf{Template 2 (\texttt{ComparedWith})}: This constraint accepts two state fields (X and Y). Based on this constraint, a new state is generated when the system detects a change in the  relationship between X and Y. For instance, if X and Y are \textit{speed} and \textit{takeOffSpeed}, respectively, there will be three constraints generated as  \textit{speed} $<$ \textit{takeOffSpeed}, \textit{speed} $==$ \textit{takeOffSpeed}, and \textit{speed} $>$ \textit{takeOffSpeed}. Thus the tool merges all concrete states in the trace in one of these abstract states.

\textbf{Template 3 (\texttt{ComparedWith a Range})}: This constraint accepts three state fields (X, Y, and Z) and is an extension of previous template. Based on this constraint, a new state is generated when the system detects a value change of field X against the interval (Y, Z). Thus the possible states would be: X $<$ Y, X $==$ Y, Y $<$ X $<$ Z, X $==$ Z, Z $<$ X. In the example from Table \ref{constraintsTable}, \textit{altitude} is compared with  [\textit{groundAlt} , \textit{safeAltForGearRetract}]. Note that, in general, Y in the previous template and Y and Z in this one can be system variables or constants.

As discussed, the user can define one or multiple constraint(s). There can be even more than one constraint per field. 
To create an abstract trace from a concrete trace, the tool reads the concrete execution traces sequentially. 
For each trace, it checks the defined constraints against the current set of values for the involved fields to determine whether a new state should be generated in the system, or not. 
Each selected function call in the trace will either take the execution to 
a) the same state (loop), b) one of the existing states other than ``self'', or c) creates a new state and transition between the two states. 
Note that in (a) and (b) if the transition already exists it would not regenerate it.  

\textbf{Extra Template (\texttt{TraceStates only in Range})}: Another important practical feature of our tool, in the abstraction step, is to set a limit as an interval for a state field values. 
The limit is defined the same as other constraints and is called \texttt{Trace}-\texttt{States only in Range}. 
This constraint is not an abstraction mechanism, but is rather a filter that helps the extracted state machines to be practical, specially, when using Template 1 in a debugging application. 
The filter sets a limit on a field value X as a range (e.g., \textit{0} $<$ \textit{takeOffSpeed} $<$ \textit{100}). 
This results in completely ignoring parts of the trace where X is beyond the interval, and then applying the other selected constraint(s) for the actual state merging. . 
This is helpful in the cases when a frequently changing field needs to be selected under \texttt{ValueChange}. 

It is important to note that this set of Templates is tailored for our case study requirements, after discussion with the company experts.
It means that the Templates are not meant to be a comprehensive set that covers every possible scenario for every system. 
Even if we would come up with such a comprehensive list, proper evaluation of its efficiency requires multiple case studies from multiple application domains, 
which is beyond the scope of this paper, which is an experience report in a specific case study. 
However, adding a new Template or modifying existing ones in our approach is straightforward and will not affect other steps.
It basically requires domain expertise about the scenarios that can not be accurately abstracted by the current set. 

Also note that there is always an option of saving a configuration (e.g., a constraint for a field) to be used as default later on. 
This means that, over time, all fields will have a default constraint that can be loaded automatically.

\begin{figure*}
\centering
  \includegraphics[width=0.5\textwidth]{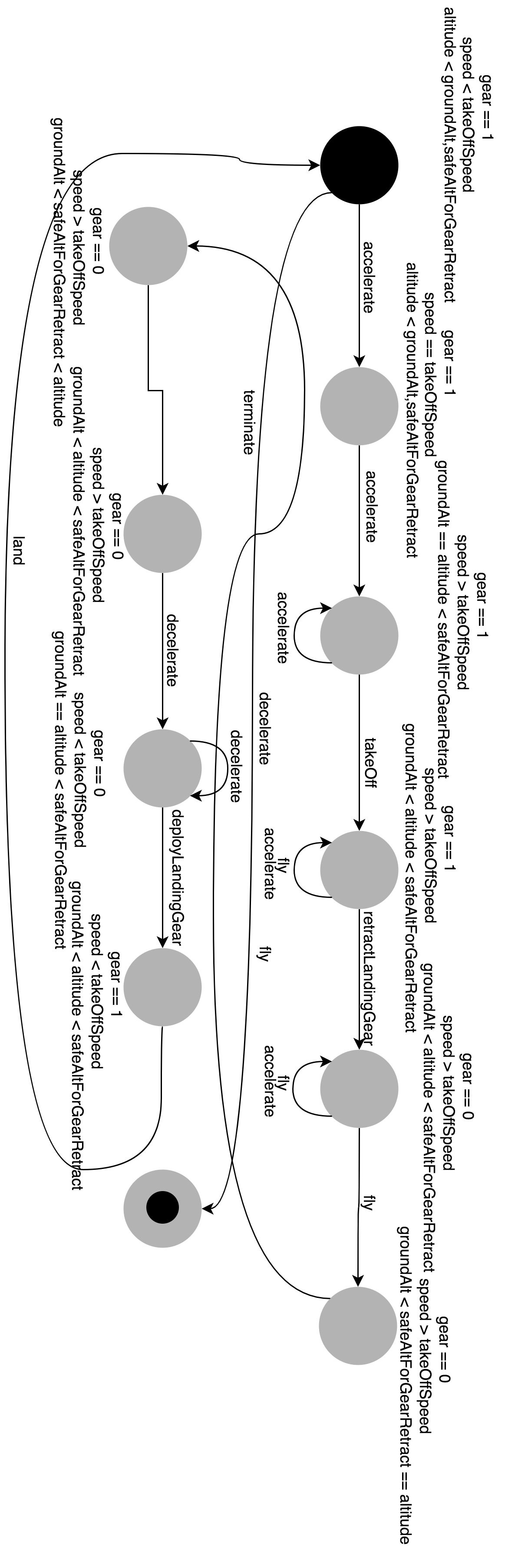}
  \caption{Example of a final state machine with combined constraints.}
  \label{figure:combinedTraces}
\end{figure*}

Finally, in \textbf{step 8}, all abstracted traces are combined into one state machine. In fact, this is not a one-step task. 
The state machine is generated gradually, when each concrete trace is abstracted and appended to the existing state machine (starting from nothing, to one path, and finally becoming a full representation of all monitored traces).  A sample state machine with the combined constraints of Table \ref{constraintsTable} can be seen in Fig \ref{figure:combinedTraces}.

\textbf{Extra Feature (\texttt{Zooming-in})}: The last practical feature of our tool in the abstraction step is allowing ``Composite States''. Basically, whenever, we abstract concrete states into higher-level states, we do not discard the low level states. The tool allows the user to select a state and zoom in. This lets the user to see all the concrete states and transitions (actual method calls) within an abstract state.  This is a perfect debugging feature where the user can zoom into the problematic states and dig into the issue with all execution-level details without being lost with a detailed full state machine. 



Finally, in \textbf{Step 9} of the Fig \ref{figure:functionality}, the user is presented with the final abstracted state machine in an interactive web form, where the user learns the context of the executed scenarios. The models look like basic Finite State Machine, except that there are state invariants on each state, which are defined using the template constraints on the system variables. The user can interact with the web form to either change the constraints in Step 7 to regenerate the state machine with different abstraction level, and/or modifies the list of monitored fields and functions to fine tune the context. When the issue is localized in the state machine the user digs into the problematic states (by zooming-in to the problematic states) to see the execution-level details and debug the issue.

\section{Empirical Study} \label{empiricalstudy}
In this section, we will explain the details of applying our approach to an industrial case study and report the results.
\subsection{Objective of the Study}
The goal of this study is to investigate the feasibility and usefulness of our interactive specification mining approach for debugging, in a real-world industrial setting. To achieve this goal we have broken it down into four research questions, as follows:

\textbf{RQ1)} Does our interactive specification mining approach correctly extract the behaviour of the running system?

This question verifies that our approach, in general, is sound. Since there is no correct state machine already in place in our context, we verify the outputs models by inspection, using developers knowledge. 

\textbf{RQ2)} Is selecting the relevant fields, functions, and constraints, interactively by the developers, feasible?

This is the core assumption of our approach that the semi-automation is feasible in terms of overhead and complexity, in this context. Unlike most existing related work in specification mining domain, we do not try to fully automate the abstraction process and will get the key insights from the domain expert. Therefore, the question is whether the domain expert is able to provide those inputs, in practice, and how costly it is for them to provide such inputs? 

\textbf{RQ3)} Does the abstracted behaviour provides any extra useful insight compared to the current state of practice, for debugging?  

Assuming our approach is sound (RQ1) and feasible (RQ2), the next question is whether its provide any extra insight for debugging that makes it useful in practice.

\textbf{RQ4)} How easy is it to adopt this approach, in practice?

In RQ4, we briefly investigate the applicability and usability of the approach, in practice. This includes challenges of incorporating this tool into the company's day-to-day toolset. 

\textbf{RQ5)}  How effective is our semi-automated approach in comparison with traditional fully-automated techniques?

Finally, in RQ5, we will quantitatively compare our approach with some well-known model inference techniques. We will investigate the alternative techniques' feasibilty as well as effectiveness.   

\subsection{Study Context}
We conducted our study on a large code-base of a safety critical embedded system (an autopilot software for UAVs), developed and owned by MicroPilot Inc. The Autopilot is a huge code-base with over 500 KLOC in hundreds of C/C++ source files. 

The company is interested in acquiring safety certification (DO-178C) for its Autopilot software, where one of the main requirements is to have explicit specification of the system, with traceability to source code and test cases, and vice versa. The company is also interested in increasing their code coverage of unit test cases, and providing a better tool support for monitoring and debugging.

Our research project \citep{micropilotProject} with the company has started in 2015 and covers several of these challenges, including a model-based system test generation, a search-based unit test generation, and the current work on specification mining. Though the specification mining approach is useful for other purposes in the context of the project, in this paper, we focus solely on its application for debugging, which was motivated by our industrial partner, as a priority. 

\subsection{Experiment Design}
Our experiment design consists of a qualitative (interview) study to cover RQs 1--4 and a quantitative comparison to answer RQ5. In the rest of this section, we will explain our design details.  

\subsubsection{Interviews}
The qualitative study was designed by interviewing the developers employed by our industrial partner, which will potentially be the actual users of the tool. A total of three rounds of interviews were conducted at the company premises. During which, the developers were allowed access to the source code, a set of test cases, and issues. The test cases and issues were selected from their bug tracking system. All the issues used were encountered in the history of the system from a year before the experiment to the time of experiment. The details of the interview process are explained in the following subsections.

\subsubsection{Subjects of Interviews}
\label{InterviewSubject}
A total of eight employees (subjects) were interviewed in this study, in three rounds. The self-reported demographics is summarized as follows:

Six out of eight interviewees are software developers, one team leader and one working as a control engineer. One of the developers has a Ph.D., three of them have Masters degree in Computer Science, while the remaining developers have Bachelors degree in either Computer Science or Electrical and Computer Engineering. The total experience of six out of eight developers in the software industry ranged from one to three years. One developer had five years of experience while the team lead had 13 years of industrial experience, all with the same company. The subjects' level of familiarity with the software under study (Autopilot) ranged from one to three out of five (self-evaluated), with the exception of the team leader who had a familiarity level of five.

A total number of five developers were included in the first round of the interview. The second round, which was a feedback validation step, consisted of only two developers from the first round. The third and final round of the interviews was targeted towards the actual use of the tool where all eight subjects were used; including the Ph.D. holder and the team lead.

\subsubsection{Pre-interview Tutorial}
Three weeks prior to the first interview, we did a quick survey of the developers' knowledge on UML, State Machines, and Software Modeling, where they were asked whether they are familiar with those concepts. Four  subjects were familiar with the state machines, out of which two had seen them only during their academic studies. Therefore, to refresh their background and making sure they are all at the same level, we held a short tutorial on state machines, abstraction, and specification mining in the company, where we not only briefed them on general ideas of abstraction and state machines but also demonstrated our tool. 

The tutorial was designed and run by the second author of this paper and one industry staff (a senior developer who was our main technical contact person at the company and helped us in designing the interviews, but was not among the subject groups).

During the tutorial, the subjects were shown a few example state machines depicting the general behavior of the autopilot during certain aspects of flight (e.g., takeoff, landing, etc). While showing them the state machines, the functionality of the tool was also explained, including a very brief (5-10 min) training of how to read the output state machines, view information related to a certain part of the execution, and relate the flow of the state machine to their knowledge of the system. They were also asked if they think the behavior of Autopilot shown in the sample state machines looks correct or not, where all subjects verified the correctness.

\subsubsection{Issues}

\begin{figure}
    \centering
    \includegraphics[width=.9\columnwidth]{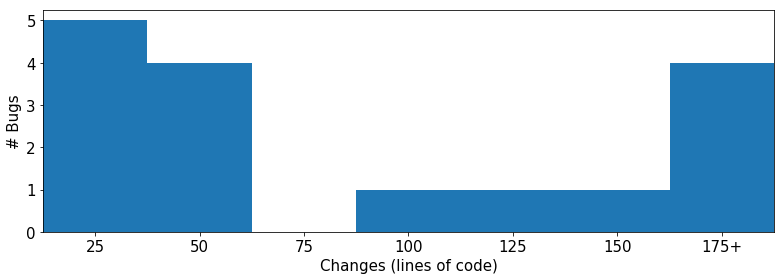}  
    \caption{\label{fig:changed_lines}  Distribution of patch sizes (in terms of lines of code), per bug.}
\end{figure}

For the actual interviews, we used a pool of 17 real reported issues. 
Due to confidentiality reasons we can not reveal the issues themselves, not even an anonymized version.
Each issue contains a description of the program behaviour and the expected behaviour as well as a summary in the form of
a title. Some included discussions of the developers about the issue. So the issues were in the form of wrong behaviours
of the autopilot software in certain situations.

The issues were selected with the help of a senior developer
from the recent history (less than a year old) of the company's issue tracking system, 
to be representative of their Autopilot's typical issues. 

All these issues were already resolved and closed. Since a version control system is used in the company and 
they mention the issue id in commit messages that are related to fixing bugs we could revert the code back
to the old buggy version and reproduce each issue in a separate copy of the code-base. We made 17 buggy versions of the code
where each buggy version has only one known unresolved issue. 
All these buggy versions were validated by our business expert (technical contact person at the company who was not an interviewee).

The main criteria for selecting an issue were 
a) the bug should be realistically reproducible in the most recent stable version of the code 
(e.g., the older versions of the code were not executable because of licensing changes of internal tool sets), 
b) the subjects must not be involved in resolving the issue, and 
c) they should be representative of typical issues encountered in everyday activities of the developers in the company. 

These criteria were validated by the business expert. 
Point (b) was also double-checked against the history of code changes and involvement of the subjects on the selected issues and fixes.

The standard procedure at the company for debugging is: a) reading the bug report, b) writing a test to reproduce the bug, c)locating the bug and debug it, and d) add the test case in their regression tester framework. Since the final test case includes assertions that were defined only after debugging, to emulate a normal debug scenario, the abstraction tool uses the test case to reproduce the bug but does not use it's assertion data.



    

\begin{figure}
    \centering
    \includegraphics[scale=0.4]{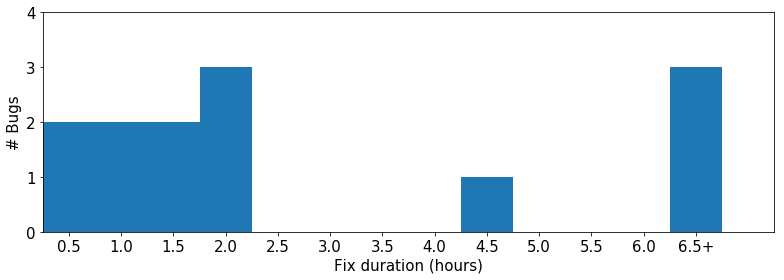}  
    \caption{\label{fig:time_taken} Distribution of historical Fix Duration (including fault localization, debug, and repair).}
\end{figure}

Since the bugs that we used in this study were already resolved, we know their fix duration (from the issue opening time to its closing time as a resolved issue) without using our tool, based on their time tracking system (Figure \ref{fig:time_taken}).  We also calculated the number of lines of code that have changed to fix the issues (patch size). These variables can be used to get a sense of these bugs' complexities. Note that we found 6 outliers in the patch size (and consequently the fix duration). Looking deeper into those cases we found out that the anomalies are attributed to several reasons: Adding new unit tests, reformatting the code, refactoring (such as renaming variables), adding new log messages, and adding new phrases to the translation tables. These factors increase the patch size and make the bug look way more complex than it is, thus we treat them as outliers.  We tried to mitigate this issue by excluding translation files and unit tests. After applying this process, patch size became more representative of the complexity of the bugs (See Figure \ref{fig:changed_lines}). However there are no means of excluding those times from the outlier issues' fix duration.

\subsubsection{Interview Design and Setup}
As mentioned, in this study, a total of three rounds of interviews were conducted.
The second round was held one month after the first round and 
the third round was done two weeks after the second round. 
The purpose of Round 1 was to get the initial feedback on the approach explained in Chapter \ref{methodology}. 
Round 2 was performed as a follow up to verify the changes we made to address some of the feedbacks from Round 1.  
The last round was performed so that the users can use the final tool within a realistic context and provide their feedback.

\paragraph{Round 1:}
\label{round1}

The first round of interviews was designed to assess the ability of our tool to abstract the behaviour of a buggy scenario to help for debugging. In other word, in this phase we only want to know if a) an appropriate model can be inferred in practice from buggy traces, using specification mining techniques and b) whether there is a potential for those models to be useful for debugging. 

To evaluate that, each subject were provided with handouts containing description of five random issues/bugs from our pool of 17 issues.
As discussed, the descriptions did not reveal the fix. 
Each issue came with its corresponding faulty and correct code, before and after the bug fix. 

Since the objective of this round was to verify the generated models and evaluate their usefulness for debugging (not evaluating the feasibility of the semi-automated approach and its overhead cost), 
the correct and buggy models were already generated (from the buggy and fixed code) and given to the subjects. 

The models were created prior to the interview by the authors, from a set of fields, functions, and constraints given by the domain expert.
In other words, round 1 does not evaluate the difficulty of selecting fields, functions, and constraints. These models and the ones created by the interviewees in the next rounds were all EFSMs with number of states in the range of 3 - 12.

In this round, we were given a budget of 60 minutes, per subject. 
We divided it to six 10 minutes intervals (10 minutes per issue plus a last 10 minutes for answering a questionnaire in a one-to-one interview, which was recorded and transcribed later).
Since the full debugging process (including fixing the issue) could be quite time consuming (according to our business expert might take up to several hours or even days), 
we did not ask the subjects to fix the bug in their 10 minutes, but rather inspect the models and see if they can spot the issue.

As part of the model validation procedure, the subjects were also provided with the list of selected fields and functions in the handouts. 
They were also asked if they would have selected the same fields for the issues. 
Note that the actual evaluation of fields and function selection is done in Round 3.

In each interview session, the subjects were asked to 
(a) check if they can relate the faulty and correct behavior in the state machines to the given issue descriptions, 
(b) explain the differences in behavior in the two state machines, 
(c) identify the fields relevant to the issue, 
(d) locate the buggy function, and 
(e) answer some feedback questions about feasibility of the overall approach, effectiveness of state machines in finding bugs in general and usability of the tool.

It is worth mentioning that as we discussed in the related work section there exist techniques that can automatically detect differences between two or more state machines. Therefore, for debugging scenarios (like regression testing) that there are two models to compare, one can actually do this more automatically. However, this was not part of our Interview objectives, as explained.

The detailed questions and the summary of the answers will be provided later in this section as well as in the Appendix section.

\paragraph{Round 2:}

The second round of interviews was a follow-up to the first round. 
It was performed to get feedback on the newly added features that were requested by the developers in round 1. 
The two developers from the previous round that asked for the new features, participated in this round.

Due to time constraints the number of issues per subject were limited to two. 
For the first issue, the interviewees were provided an already selected set of fields and functions, and constraints. 
Hence, they just had to generate the state machines.  
For each issue, they were required to use the tool twice to get the buggy and the correct state machine and inspect them similar to round 1.

For the second issue, they had to select everything by themselves. 
The subjects were not provided with the models, but rather asked to use the actual tool to create the models, by themselves. 
The second issue of round 2 and the entire round 3 were carried out to show the cost-benefit and feasibility of our approach by asking the interviewees to select the inputs and constraints.  

In addition, unlike round 1 and round 2-issue 1, where the task is just spotting the difference between two state machines, here (round 2-issue 2) we ask the subjects to identify the location of the bug, without the correct version. 

\paragraph{Round 3:}

The objective of this round is to extend round 2-issue 2 interview with more bugs and more subjects.
In addition, we want to properly compare the benefits of behaviour abstraction compared to the company's current debugging practice. 
It involved two mid-level developers and a team lead who had not seen the tool before the tutorial session. 
A total of five buggy versions of the code base were prepared. 
Developers were asked to choose any two of them (after confirming from the bug tracking system that none of the developers were involved in the resolution of any of the five bugs). 
For one issue, they were asked to find the location of the bug by their standard practice of debugging at the company. 
For the other issue, they were asked to use our tool for the same purpose. 

For each bug, the allotted time was half an hour. 
After the first half an hour, when the manual debugging period finished, the author provided a very brief demonstration of the tool, similar to what was demonstrated in round 1.  
During this demo the functionality of the tool, using a general scenario of Takeoff aspect of the flight, was explained. 
The tool was demoed for one issue, out of three, that were not selected by the developer. 
In the final half an hour, the subjects were asked to use the tool by themselves for debugging the selected issue. 

During the sessions, the author observed their manual debugging practice and categorized it under the ``Bug Diagnostic Strategies'' explained in \citep{dbgbench}. 
In addition, the number of executions performed by the interviewee, manually or by the tool, was recorded. 
Finally, the time taken by each execution during each session was recorded.  
At the end of each interview, they were asked general feedback questions regarding the tool and the approach. 
The results from the experiments are reported in Section \ref{interviewresults}.

\subsubsection{Interview Questions}
All three rounds of interviews were designed in collaboration with the company team (including a high-level manager), 
such that not only they answer our research questions but also assess the relevance of the questions to the company's needs 
and make sure they do not violate the employees rights and privacy.

For the first round, the final questionnaire includes three types of questions: 
a) Demographic questions (included in all rounds of interviews), where their answers are already discussed in the Subject of Interviews in Section \ref{InterviewSubject}, 
b) Questions specific to debugging an issue, and 
c) Questions about the overall idea of the tool and general feedback on its feasibility and usability.

The detailed lists of above questions are provided in Tables \ref{onedemographic}, \ref{oneinterview} and \ref{onefeedback} in the Appendix section.

For the second round, the questionnaire included questions regarding 
a) difficulty of fields and function selection, 
b) the overhead cost of model generation using our tool, and 
c) potential benefits of the newly added features (automated loading of fields when a function is selected or vice versa). 
The questions are provided in Table \ref{twofeedback} in the Appendix section.

Finally, in Round 3, we had a set of questions to be answered by observing the subjects and their interaction with the tool and debugging practices. 
These questions are provided in Table \ref{threeobservational}, in the Appendix section.
We also had a set of feedback questions provided in Table \ref{threefeedback} in the Appendix section.

\subsubsection{Comparison Study}
\label{ComparisonStudy}

To answer RQ5, we compare our approach with some of the well-known specification mining techniques implemented in the MINT framework \cite{Walkinshaw2016}. MINT is a promising tool that has implemented a number of state machine inference algorithms (e.g., kTails\cite{ktails}, gkTail \cite{gk-tail}, and EDSM \cite{langbluefringe}), in a modular way, i.e. one can easily replace components of the algorithms. Therefore, it is a perfect tool for our comparison study. 

Before using MINT, we had to go through a prepossessing step where the execution traces were converted into the MINT format.
We run MINT under various configurations, as follows:
\begin{itemize}
    \item \textbf{strategy}: This option specifies the employed inference algorithm. To be more precise, it only specifies the state merging strategy, so for example, you can use ``gkTail'' strategy but instead of using Daikon to infer transition guards, you can use another arbitrary algorithm. The possible values for this option are {\verb ktails }, {\verb gktail }, {\verb redblue }, {\verb noloops }, and {\verb exhaustive }.
    \item \textbf{k}: This is the $k$ parameter used in ``kTails'' and ``gkTail'' algorithms. We used 0, 1, and 2 for this parameter.
    \item \textbf{carefulDet}: This is also a boolean value, where having it ``on'' instructs the tool to make deterministic automatons, which prevents over generalization. 
\end{itemize}

Overall, we have tested a total of 30 (5*3*2) configurations for each bug.  More details about the options can be found in MINT report \cite{Walkinshaw2016}.

To compare debug assisting tools a number of measures exist. Xia et. al. and Xie et. al. used debug success rate and debugging time as their measurement values \cite{Xia2017, xie2016revisit}. In the survey by Parmar et. al. EXAM score is mentioned as the number of the statements that need to be examined before the faulty statement is found\cite{parmar2016}. Given that this score would not work on the model-level, we slightly modified it as the number of states that need to be examined before the faulty state is found.

\subsection{Experiment Results: Answers to RQs}
\label{interviewresults}

To answer our research questions, in this section, we will provide a summary of the answers to all questions by the eight subjects of the interviews. 
In addition, Table \ref{table:results}, \ref{table:results2}, \ref{table:results3} and \ref{table:results4}, in the Appendix Section, summarize all answers, per question/interviewee. 
The numbering of the interview questions follows the same format from the tables in the Appendix section.

\subsubsection{RQ1: Does our interactive specification mining approach correctly extract the behaviour of the running system?}
To answer this question, we asked four basic questions in Round 1 of interviews (Q1.2.1, Q1.2.2, Q1.2.3, and Q1.3.2), with the following answers:

\textbf{Q1.2.1: } \textit{Explain the differences that you notice between the buggy and correct state machines (e.g., number of states, sequence of functions, fields changes, etc.)}

\textbf{Answer: } Three out of five subjects were easily able to relate the descriptions in the issue to the behaviour illustrated in the state machines, without our help. 
They also stated that the behavior in the correct state machines conforms with the normal behavior of the program according to their understanding. 
Two subjects were initially needed guidance to go through the execution flow in the state machine, for the first issue. 
However, they managed to analyze the remaining issues without any help. 

\textbf{Q1.2.2: } \textit{After comparing both state machines, are you able to identify the field(s) that caused the issue or showed the effect of the issue?}

\textbf{Answer: } All subjects were able to identify the fields that were affected by each issue, from the list of all selected fields in the state machine. 
The purpose of this question was to verify if the subjects would be able to filter the fields, from a bigger list of selected fields, relevant to each issue.

\textbf{Q1.2.3: } \textit{Are you able to identify which function caused the issue?}

\textbf{Answer: } Since the subjects were able to identify the differences in behavior showed in the state machines, they were able to identify the functions causing abnormal behavior. 
However, the actual function causing the issue in 2 state machines were not selected, intentionally. 
For those state machines, three out of five subjects stated that they don't think the function causing the error is included in the state machines. 
They also correctly stated that the issue causing functions were called within one of the selected functions. 

\textbf{Q1.3.2: } \textit{With respect to the selected fields and functions, are the state machines generated by the system correct? Are they detailed enough? If not, what important information is missing?}

\textbf{Answer: } Answering the question, three subjects said that they think the state machines were complete according to their understanding of the system, 
while two correctly commented that some models look complete but they missed useful states that are to do with functions that are not selected in the first place, which was an intentional choice by us. 
A subject commented that sometimes selecting proper functions is difficult in the first try, since you may create a correct model but miss important details. 

A suggestion was also made to include all the important functions, which is impractical due to performance degradation and very large output state machines. 
One interesting suggestion was to warn the user if the system notices a change in the fields without any transitions (selected function calls) between them 
so that the user can analyze the change and select the relevant functions and repeat the step. 
Another suggestion was adding a new feature to perform more complex operators on the fields within the state machine, e.g., RADIUS(field1)$>$field2, which can be part of future work.  

Note that all answers are based on Round 1 interviews, where the fields and functions lists are already given.
However, that is not an issue since the RQ1's objective is not to assess the feasibility of the approach (which are taken care of in the next questions).
Therefore, given these answers, we conclude that the extracted models are correctly representing the run-time behaviour.   


\subsubsection{RQ2: Is selecting the relevant fields, functions, and constraints, interactively by the developers, feasible?}
To answer this question we asked questions both regarding the complexity of the task (Q1.3.1, Q1.3.3, Q.2.1, Q2.2, Q3.2.1, Q3.2.2) and its overhead cost (Q2.3, Q2.4, Q3.1.1, Q3.1.2) in all three rounds of interviews, with the following answers:

\textbf{Q1.3.1: } \textit{After looking at the list of functions and fields selected to a given issue, how difficult do you think it would be if you had to select those inputs for a given issue?}
Note that the purpose of this question is not to properly evaluate the difficulty of fields and function selection, which will be evaluated in round 2 and round 3, 
but we rather wanted to get an initial feedback from the subjects and collect ideas such as the predefined sets, mentioned above. 

\textbf{Answer: } Four out of five subjects graded the difficulty of the process of selecting the relevant fields and functions with respect to any given issues as two out of five; 
one being the easiest and five difficult, whereas one subject commented that if you are not familiar with the code-base, 
it will be difficult to select the exact function but a keyword search can give an idea of where to start. 
However, the subject also commented that it is even more difficult to use Visual Studio for debugging the code if you are not familiar with the code-base, and it becomes a guessing game. 
Another interesting suggestion made by the same subject was to prepare a subset of fields and functions for every aspect of flight (or any other operation) 
such that if users want to debug an issue related to, say ``landing'', they would just have to select the aspect rather than a huge list of fields and functions.

\textbf{Q1.3.3: } \textit{For given issues, are the set of selected fields and functions adequate? If you were selecting the fields by yourself, would you have made any changes to the list of selected fields and functions?}
Again, this question is a preliminary assessment and more detailed investigation of the ``selection phase'' was left for next rounds.

\textbf{Answer: } Three subjects stated that they would have selected additional fields and functions for some of the issues and would omit the irrelevant fields. 
Two subjects thought that the selected fields were adequate for them to find the issue. 
But all of the subjects were able to identify the issue from the state machines. 
Therefore, we can summarize that even when all the relevant fields are not selected, subjects can still get an idea where something is getting wrong. 
In addition, since the process is interactive, users can adjust their selection and try different inputs and see the results, immediately.

\textbf{Q2.1: } \textit{How difficult was it to select the functions / fields, on a scale of one to ten, for a given issue?}
 Note that the answers to Q2.1 and Q2.2 are based on Round2-issue2, where the interviewees had to select the fields, functions, and constraints by themselves.  
 
\textbf{Answer: } Answering Q2.1, both subjects gave the procedure of selecting fields and functions a difficulty rating of two out of ten; with one being the easiest and ten the hardest. 
They stated that the keyword search is very effective in exploring the relevant fields and functions of an issue. 
The subjects stated that being the developer of the system, this is the most basic information we can get from them. 
And even if a developer might not be familiar with all fields and functions, 
(s)he can guess their names since (s)he is familiar with naming standards followed in the source code for defining fields and functions.

\textbf{Q2.2: }\textit{After selecting the list of fields and functions, on a scale of one to ten, how difficult was it to come up with the valid constraints with respect to the  scenario?}

\textbf{Answer: } Answering Q2.2, one of the developers graded the process of \textit{selecting fields under different templates of constraints} a difficulty rating of one 
while the other graded two; 
with one being the easiest and ten hardest. 
The developers stated that since the fields are already selected by them, they have a good idea of which constraints and templates would be useful.

Finally, Q3.2.1 and Q3.2.2 extend the Round2-issue2 setup with more interviewees and issues. Therefore, we can get a difficulty scale. 

\textbf{Q3.2.1: }\textit{How difficult was it to select the functions / fields in general on a scale of 1 to 10?}

\textbf{Answer: } Answering to the question, one of the developers graded the process of selection a difficulty level of one, while the remaining two graded it three out of ten.

\textbf{Q3.2.2: }\textit{After selecting the list of fields and functions, on a scale of 1 to 10, how difficult was it to come up with the valid constraints with respect to the scenario?}

\textbf{Answer: } The grade (one) was selected by all three developers, for this question.

Based on the above answers, we conclude that the main idea of selecting fields and functions is feasible (complexity-wise), in the context of our case study, since all subjects graded the process of selection of fields and functions in a range of 1 to 3, and graded the feasibility of the approach in the range of seven to ten out of ten. 
Note that these answers are from subjects who speculated the difficulties in Round 1, did a fields/function selection from a predefined list in Round 2,  as well as those who did select the fields and functions completely from scratch in Round 3.

To get a sense of the overhead cost of our tool we asked the following questions, in Round2:

\textbf{Q2.3: }\textit{How many times, on an average, did you have to repeat the process (reselect the fields and functions, reselect constraints, and regenerate the state machine) to  abstract  the program behavior  according  to  your  need  and  expectation?}

\textbf{Answer: } Both developers used the tool only once for each configuration, per issue; once for buggy version and once for the correct version as they were satisfied by the correctness of state machines and could see the symptoms of the issue in the first run.

\textbf{Q2.4: }\textit{How much time did each run take on an average?}

\textbf{Answer: } The entire selection of fields and functions took less than a minute. 
Similarly, the process of defining constraints took less than a minute, where one of the issues selected by the developers required as many as 5 constraints in one definition. 
Basically, the compilation and execution of the source code took the most time, which is about 5 minutes, 
which confirms the fact that invariant detection techniques that require many executions to build a constraint are not practical in this context. 
The generation for both issues took less than a minute, and the subjects were able to spot the issue in under two minutes. 
Hence, the average total time per issue was less than 10 minutes.

We also observed the cost of interacting with our tool and reported them in the following observational questions, in Round 3:

\textbf{Q 3.1.1: }\textit{How many times, on average, did the developers have to repeat the process (reselect the fields and functions, reselect constraints, and regenerate the state machine) to abstract the program behavior according to their needs and expectations?}

\textbf{Answer: } All developers got the expected state machine, that showed their expected behavior, on the first time. 
Hence they did not repeat the process. 
For demonstration purposes, as it took around 10 minutes for the team lead to debug one issue with the tool, 
he was asked to try another issue. 
However, he managed to get his expected behaviour again with only one try.

\textbf{Q 3.1.2: }\textit{How much time did the manual debugging take, and how much time was taken by each tool run, (average is reported for multiple runs)?}

\textbf{Answer: }  It took anything between 10 to 30 minutes by the subjects to manually debug the issue  
(mostly including code review and occasional partial executions). 
However, the time taken by the same developers for using the tool ranged from 10 to 12 minutes.
Note that this time also includes the full compilation and execution time of the test cases, which ranged from 4 minutes to 8 minutes per execution.

Therefore, we also conclude that the cost of our approach is reasonable which makes the overall idea quite feasible in an industry setting. 


\subsubsection{RQ3: Does the abstracted behaviour provides any extra useful insight compared to the current state of practice, for debugging?}
To answer this RQ, we asked questions (Q1.3.4, Q1.3.5, Q3.1.3, Q3.2.3, and Q3.2.4) about the current practice of debugging at the company (as our baseline) and the potential advantages/disadvantages of the proposed tool, with the following answers:

In Round 1 we asked the following two questions:

\textbf{Q1.3.4: } \textit{What is your current practice/procedure of debugging the system, when an issue is reported and assigned to you? Which tool support you have?}

\textbf{Answer: } Answering the question, most interviewees responded that when an issue is assigned to them, 
they first try to reproduce the issue, as explained in the description, and run it on the simulator in their in-house tool. 
Then they move to Visual Studio, go through the code and use breakpoints to break the executions in different functions, 
then they step through the code while inspecting the values of certain fields.
In round 3, we observe these practices, in details. 

\textbf{Q1.3.5: } \textit{Assume you select the best set of fields and functions for a certain issue and use our tool to generate the corresponding state machines. 
Do you think our tool would bring any advantage to your current set of debugging/monitoring tools, at Micropilot Inc.? 
If so what advantages and if not why?}

\textbf{Answer: } Answering the question, three out of five subjects said the ``visual representation'' of the code made them see the big picture of the execution, 
they were able to see where behaviour was branching and they learned something. 
In addition, it provided them an organized way to inspect the fields during the execution of the whole system 
and gave an idea on which functions to review and in which level of detail. 

Two subjects also added that even incomplete state machines (with missing fields and functions or inappropriate level of details) are helpful 
in that they would have given them a starting point to debug (finding which functions to put breakpoints at). 
In addition, they mentioned that the stack traces that are accessible in the detail (zoom-in) views are also helpful for adding the caller functions into the selected function list. 

One subject also added that it is easier to explain the functionality of the system with respect to fields and function calls on the state machine 
compared to explaining the source code (i.e., the program comprehension application of abstraction).

Another subject stated that ``if a relevant field is changing too many times, it's hard to debug the function that is changing it. 
You might miss the important changes that you are actually interested in when skipping breakpoints. 
So, it's interesting to see how the field changes throughout the execution in the state machine and then go back to the code to inspect in relevant intervals.''

Next, during Round 3, the following observations regarding their diagnostic strategies were made, while the subjects were working on their tasks. 

\textbf{Q 3.1.3: }\textit{Which diagnostic strategies were used?}

\textbf{Answer: } The three developers used a mix of three debugging strategies: Forward Reasoning, Backward Reasoning, and Code Comprehension, as defined in \citep{dbgbench}. 
In forward reasoning, the developer starts from the starting point of the execution and move towards the first occurrence of the issue in search of the cause of the issue, while executing the source code. 
However, in backward reasoning, the developer starts from the first occurrence of the issue and moves backwards to the start of execution while searching for the cause of the issue. 
In code comprehension, developer reads through the code while making an understanding of the source code and try to find the issue with his mental picture of the code.

Finally, we asked the following specific questions regarding RQ3, in Round 3:

\textbf{Q3.2.3: }\textit{What is your current practice/procedure of debugging, when an issue is reported and assigned to you? Which tool support do you have?}

\textbf{Answer: } The answer to the question is divided into two parts: a) our observation on their practice during the manual debugging session and b) their self-reported practice of typical debugging. 
From our observation, one of the developers used ``Backward Reasoning'', by first reaching to the point where the issue was first noticed and then step-by-step moving back to the location of the issue's root cause. 
While the other two used Code Comprehension and Forward Reasoning to reach the location of the bug.

From their responses, it was found out that more experienced developers typically use the debugging strategies of Input Manipulation and Offline Analysis, 
while the developers with entry to mid-level experience use a mix of Forward Reasoning, Backward Reasoning and Code Comprehension strategies, when debugging an issue. 
In the input manipulation technique, the developer keeps modifying the input that is producing the wrong result and compare with the expected output, until he figures out the relation between input and output. 
In offline analysis, the developer relies on the post execution data such as log and execution traces to debug the issue.

In terms of tool support, the majority would use Visual Studio and its debugging features as the company's base IDE.

\textbf{Q3.2.4: }\textit{How beneficial is using this tool compared to your current practice?}

\textbf{Answer: }Since the current round of experiment involved developers with more experience than of the previous rounds, we received more concrete responses to this question than those in Round 1. 
The most senior developer said that the tool is useful for analyzing the behavior of the system and also for verifying the modification of the system in subsequent releases. 
The tasks can be delegated to the developers who can verify their own changes by this tool. 
However, (s)he thought that for debugging purposes, the tool is useful for entry to mid-level developers but may not be useful for highly experienced developers due to their familiarity with the code base over the tool (and state machines in general). 
One developer states that it seems useful as it can easily point the user to the actual location of the issue (function) to start their further investigation, and then come back to the tool for a more detailed state machine defined on the basis of the investigation results. 
The last developer stated that the tool looks very beneficial in situations where Backward Reasoning and Forward Reasoning are not possible due to the unknown effects of the issue on the system or for more complex issues.
He said that although as opposed to manual debugging it needs a full execution, which takes more time, in most situations the tool would take the developer to the location of the issue quicker than manual debugging practices.


\subsubsection{RQ4: How easy is it to adopt this approach, in practice?}
Finally, to answer RQ4, we asked two similar questions in Round 1 and 3 (Q1.3.6 and Q3.2.5) about the about the potentials of the tool to be adopted in the company, with the following answers:

\textbf{Q1.3.6: } \textit{How easy it is to embed this tool into your current frameworks and infrastructure? 
Any challenges that you see in making this as part of your tool set?}

\textbf{Answer: } Answering the question, all subjects agreed that the tool would be a valuable addition to their tool-set and will be worthy to adapt to the tool. 
Similarly, all subjects thought that the idea was easy enough to be applied and thus is a valuable addition.

Three subjects said that if the tool was included in their tool-set, they would adapt to it, easily. 
While two of them mentioned that any new tool is hard to adapt, in the beginning. 
However, once they get used to it, it would not be hard to use.

\textbf{Q3.2.5: }\textit{How easy is it to embed this tool into your current frameworks and infrastructure? Any challenges that you see in making this a part of your tool-set?}

\textbf{Answer: }In response to the question, the team leader stated that the biggest challenge in such tools is the maintenance, which is costly since the actual developers of such research tools move on. 
The other two developers stated that with any new tool there is a learning curve but they see no major issue adapting to this tool, as a user, once included in their tool-set.

Finally, at the end of Round 1 and 3, we ask the interviewees to comment on feasibility, effectiveness and usability, which is summarized here:

\textbf{Q1.3.7: } \textit{Please provide your feedback about the feasibility, effectiveness, and usability of this tool and idea.}

\textbf{Answer: } The subjects graded (out of 10) feasibility, effectiveness, and usefulness of the idea and the tool and provided free feedback about the whole idea of the project.

The subjects graded the feasibility in the range of 7 to 9, effectiveness in the range of 5 to 9 and usability in the range of 7 to 9. 
The grade 5 in effectiveness was given by one subject (Subject 5), where other feedback were in the range of 7 to 9. 
The grade 5 was given because Subject 5 thought that for a subject not much familiar with the system it would be harder to use it effectively. 
However, the subjects with more experience and more familiarity with the system (Subject 1 and 3), thought they will be able to use it effectively and graded effectiveness as 9 and 8, respectively. 

All the subjects think the tool is usable and the steps from selecting the fields to generating state machines were minimal and seemed easy enough
 and they would be able to use the tool without much help. 
 A subject, also, suggested that the tool could automatically compare models and flag differences, which is in our future work.

\textbf{Q3.2.6: }\textit{Please provide your feedback about the feasibility, effectiveness, and usability of this tool and the idea of using your knowledge to break the issues down to fields and functions.}

\textbf{Answer: }The team leader responded that (s)he needs to see more evidence about the feasibility of the tool for other issues in the system. 
However, (s)he graded the tool 8 for both effectiveness and usability. 
The other two developers graded the tool 10,9,8 and 9,8,9 for feasibility, effectiveness and usability respectively. 

To wrap up RQ4, most subjects didn't think they would face any trouble if they use the tool on regular basis. 
Those who found it difficult said the tool was worthy to learn, looking at the advantages it has to offer over traditional techniques. 
The subjects also thought that the idea is effective and the tool is usable for them.


\subsubsection{RQ5: How effective is our semi-automated approach in comparison with traditional fully-automated techniques?}
\label{quantitative}

To answer this RQ, as explained, we have designed a study to compare a total of 30 (5 different model-inference strategies with several configurations) alternatives, per bug. However, given the size of our designed study, we first randomly selected two out of 17 bugs, for a preliminary comparison. Two execution traces with 30 configurations made 60 model inference attempts. Unfortunately, MINT failed to infer any model for 75\% of those attempts. Failures happened due to several reasons. We have further categorized them into the three classes, as follows: a) running out of memory, b) running out of stack space (stack overflow), and c) running out of time.

All of these three categories of failures can be attributed to the sheer size of the execution traces that we had. The largest test data that could be found in the MINT repository has less than 1/50 the number of events in our average execution trace. 

In terms of category c, we set MINT's maximum timeout as 20 minutes which is far more than our tool's average abstraction time which was 5 minutes (including compilation, execution, and abstraction; as stated in the answer to Q2.4). However, even with four times higher budget for the abstraction phase alone, $28.3\%$ of the attempts failed due to timeout. Such a long run-time can be completely impractical in the context of debugging.

Overall, only 25\% of the attempts (15) ended up generating an output within the 20 minute time frame, successfully. In fact, all these 15 success cases were belonged to one bug and for the other bug under study no configuration could infer a model. Out of these 15 outputs some were completely identical, so we grouped the identical outputs together and realized that only 7 unique state machines were generated. Two examples of these state machines can be seen in Figures \ref{fig:mint2s} and \ref{fig:mint4s}.

As you can see, neither of the successful configurations created data related transition guards, even though we had asked for it explicitly {\verb daikon = ``true''}. That is, none of these outputs could be called an EFSM; all were simply FSMs. This is most likely to do with our problem context, where unlike typical use cases of specification mining, we do not have many successful traces, but rather a few buggy ones. Therefore, there won't be many sample usages of parameter values on the transitions to infer useful constraints from.

Nevertheless, for the context of debugging, FSMs are really limited, since debugging scenarios typically require inspecting variable values. Particularly, in the case of this bug, which MINT did create a FSM for the model could not reveal the bug.  Therefore, we do not consider these 7 state machines as successful runs.   

Given the extremely poor results on the preliminary study, we decided not to follow this path and did not attempt running the rest of the bugs with the given MINT configurations. We also tried some more adhoc configurations on some new traces, but the overall trends were the same. This means that we could not even measure our modified EXAM score on the alternatives (since there were no EFSM generated). 

So our conclusion for RQ5 is that the current fully-automated techniques are not ready to create appropriate models for debugging real bugs of large scale systems, at least, in a reasonable time. Thus our fast, and precise semi-automated method is a more promising approach, in practice. 


\begin{table}
\begin{tabularx}{\columnwidth}{XX}
    \includegraphics[width=\linewidth]{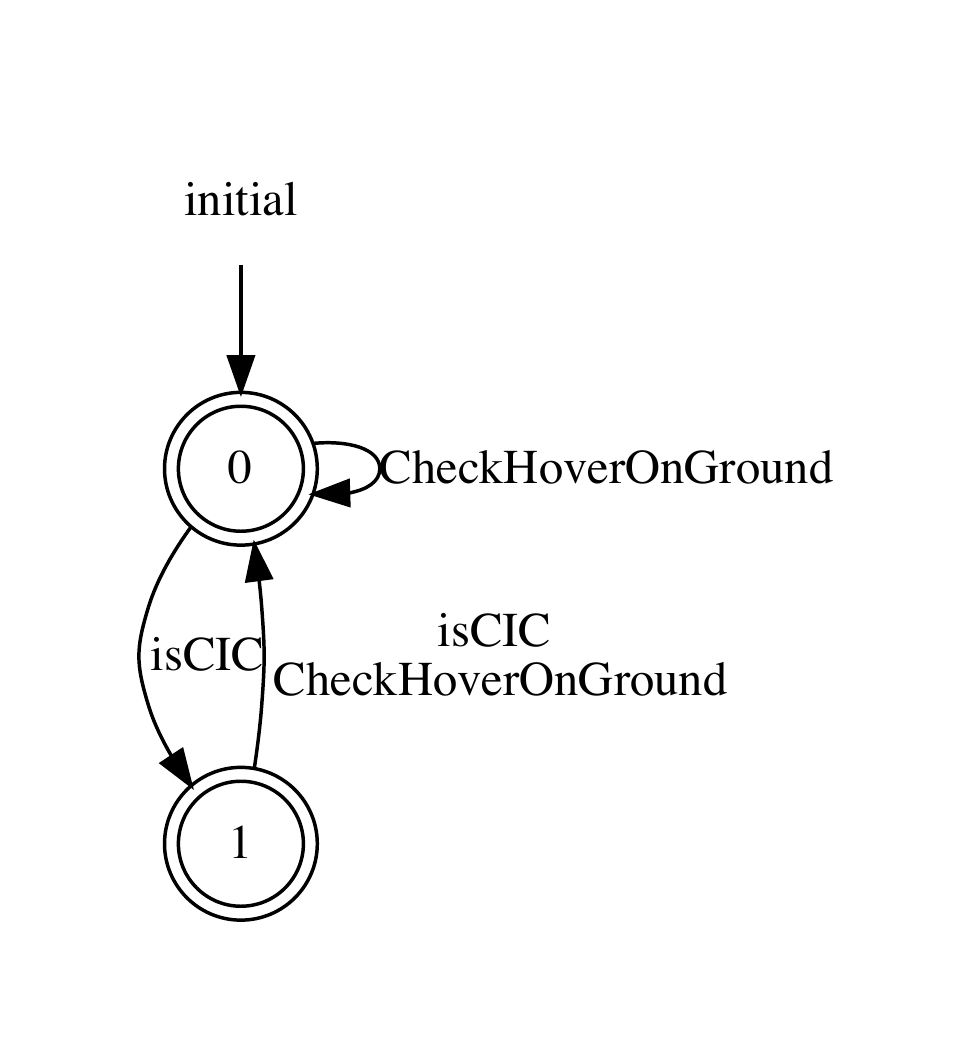}
    \captionof{figure}{A state machine generated by kTails algorithm with $k$ = 0}\label{fig:mint2s} 
    & 
    \includegraphics[width=\linewidth]{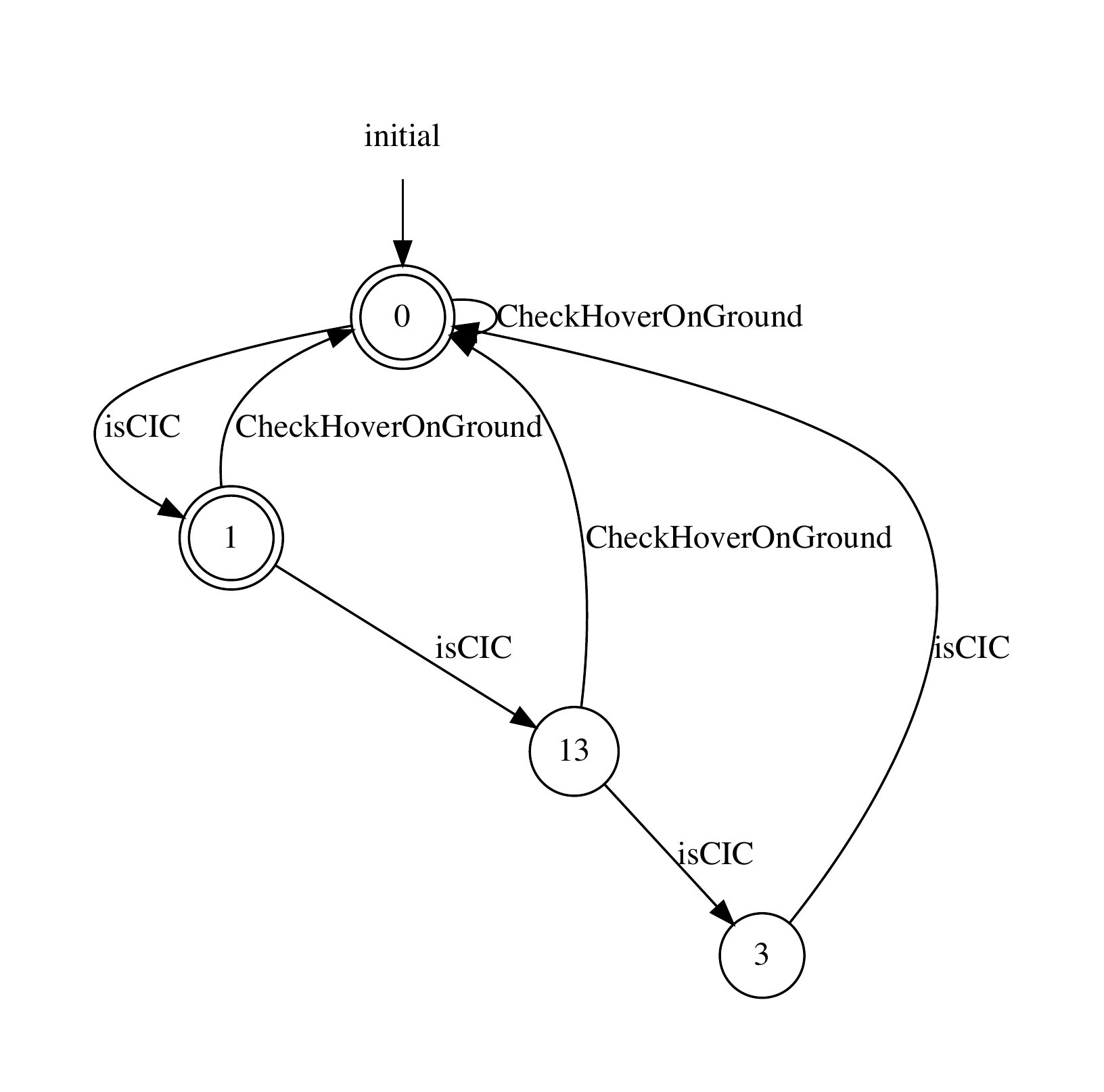}
    \captionof{figure}{A state machine generated by gkTail algorithm with the same input as figure \ref{fig:mint2s} and k = 0}\label{fig:mint4s}
\end{tabularx}

One way of quantifying the effectiveness of a model-based debugging approach is to measure the portion of the model (state machine) that needs inspection.  Since in our technique, human is kept in the loop to perform a white-box filtering, the resulting state machines are quite concise. In this case study the number of states ranged from 3 up to 12. The developers examined the whole graph at a glance. If this filtering was not done, the resulting models could have become disproportionately large. In that case a measure defined as `number of states to examine before the bug is found' could be useful. 

\end{table}

\subsection{Discussions, Threats to the Validity, and Limitations}
There are several typical threats to the validity of this study as well as limitations and critiques. 
In this section, we will explain the most important ones and our approach to tackle them.

Perhaps the main critique of a semi-automated specification mining approach is the manual fields, functions, and constraints selection process. 
There are two potential critiques here:  its cost and complexity.
To address these critiques in Round 3, we specifically observed the time taken for each debugging (broken down into different steps). 
The results showed that the manual aspects of our tool does not put an unreasonable overhead compared to the time taken for building and executing the system. 

We also asked the subjects about the difficulty of the manual parts (fields, functions, and constraints selection) and did not get a serious complaint.
Though this is a small case study and does not reflect the reality of the challenge in a bigger context, the interview results give us some confident that the interactive approach is not an out-of-question solution and must be explored more in the future. Specially, given the fact that this approach brings a cheaper (no need for several execution to extract the invariants) and more accurate solution (exact models based on user-defined inputs rather than heuristics).

As discussed in section \ref{quantitative}, other tools failed to create models to be used as the baseline for comparison. Also, EXAM score would have been the same for all of our inferred models because the models inferred using our tool were so concise (3-12 states) that the interview subjects could just get the whole picture at a glance. Therefore, a quantitative measure like EXAM score could not give much insight into the effectiveness of our model.

The tool developed throughout this research is implemented, tested, and configured by two researchers. 
There might be some bugs in the tool that are yet to be found. 
However, the output of the specification mining tool, i.e. the state machines, were constantly validated by our contact lead developer at the company. 

Another threat to the validity is being too specific to our program under analysis (Autopilot) and/or the company's debugging process. 
For example, we only monitor a certain data record (plane struct) since in our simulator all the required data fields are in there.
In addition, we do not handle multiple airplanes (defined in multiple threads) in the simulator at the current version. 
However, in general, Autopilot is similar to any other industry standard safety critical embedded systems. 
Hence, the idea can directly be applied to any other single thread embedded control software system maintaining global state fields, with very little modification. 
In addition, the company's debugging process is very typical and not specific to this company, at all. 

The pool of bugs in this study is relatively small (only 17). 
However, the bugs are real (from the history of the software) and selected with the help of our industry partner to be representative of their typical bugs. 
They can also be reasonably generalized to a broader context, 
since our approach is not designed for a specific bug but rather captures and abstracts any functional difference in the inferred models. 
The EFSMs, however, work only on the function-call level now. 
Thus the smallest unit that we can localize a bug in is the last function in a call stack that contains the bug. 

Another validity threat is to do with generalization of the results. 
Ideally, we should interview more subjects to gain more confidence.
However, note that in our case, we do not want to make the claim of generality, 
but rather that of transferability to people with the same needs. 
That's why our qualitative method is a mix of interview and observational study, which as Isenberg et. al. say \citep{Isenberg} 
can be very useful even with a ``low'' number of expert participants. 
Also note that having a larger set of participants was not easy, since they must be experts (among Micropilots developers), 
to be able to comment on the current practice of debugging vs. the use of our proposed approach. 
So we could not easily replicate it with methods such as web-based crowed-sourced interviews or student subjects (also note the confidentiality issues).
In addition, our subjects had to spend several hours (including the tutorials) with us in-person. 
Therefore, the interviews were expensive for the company. 
Thus, though, our interviewees sample size (8 developers) looks small, 
it is important to acknowledge that there is no magic number for the minimum number of participant, 
in fact, ``enough'' is the amount where additional participants don't provide any additional insights (``saturation'').
For instance, Guest et. al. \citep{guest2006many} propose that saturation often occurs around 12 participants in homogeneous groups.
In our case, when the focus is on testing the
usability/feasibilty of a tool rather than its comparative performance, there is an ongoing debate of how many participants to test. 
For instance, Nielsen and Molich \citep{nielsen1990heuristic} found that with six people one can
find, on average, about 80\% of all usability problems.
Thus we can consider the reported results, within the scope of the objectives, valid. 

As mentioned in Q 3.1.1, our interview subjects were experts in the software and were quite familiar with the code, so they happened to not need any iteration over functions and constraints selection (first time choices were good enough). While this is expected in many debugging settings, where the bugs are assigned to the experts, but the assumptions are not necessarily generalizable to all software teams. Developers who are not experts in the code they are debugging, might need to iteratively refine their inputs to get the best result out of this tool, quite similar to what they would do in a manual approach using breakpoints etc.

\section{Future Work} \label{futwork}

We are planning to extend this work 
by implementing more advanced miners from the literature and comparing their cost-benefit with the existing approach.  
We are also planning to improve the current tool by defining more templates for the abstractor and apply it on more scenarios.

We will also examine advanced techniques for automating fields and function selection.
This approach will be based on code-based fault localization techniques.
Basically, for each new bug report, we plan to learn a likelihood of a function to be buggy based on how in the past the function was part of a patch with respect to a similar reported bug.

Finally, a potential future direction for this research project is automatic comparison of behavioral models that could help even more in the debugging processes. 
For instance, for regression testing, the tool could be automated to generate and compare the state machines, automatically, 
whenever a new check-in event is triggered (in the version control system). 
This would help the users in verifying the effect of their modification after each check-in.

\section{Conclusion} \label{concl}

Fully automated specification mining techniques have been studied in literature for many years but they have not been used in industry much yet. 
One of the main reasons is that it is hard for the automated abstraction techniques to generate models in the exact level of details that the user needs. 
Moreover, invariant inference techniques that are typically used for abstraction require several execution of the system under a diverse set of scenarios to be able to accurately capture the behavioural models, which is quite costly for contexts such as debugging.

In this project, we evaluate the idea of a semi-automated specification mining, in terms of feasibility and usefulness on a real-world setting for debugging purposes. 
We proposed an interactive approach for specification mining, 
where users' domain knowledge is collected as inputs in the form of fields, functions, and constraints, 
which is a basic knowledge developers are supposed to work with on any code base. 
We implemented a tool that focuses on the debugging application of specification mining 
and let the users easily change the perspective and/or level of details 
by selecting different fields to monitor and constraints to use. 
Although developed in collaboration with Micropilot, the approach is quite generic and very light-weight.  

We evaluated the idea on a full size industry setting with real faults on the actual code-base, through a series of interviews with the company developers, 
which confirmed its feasibility and usefulness, compared to the current state of practice, in debugging.
  
\section*{Acknowledgements}
This work is partially supported by the Natural Sciences and Engineering Research Council of Canada [CRDPJ/515254-2017].
\doclicenseThis

 \bibliographystyle{elsarticle-harv.bst} 

 \bibliography{main}
 
\newpage
\section {Appendix}
\label{appendix}
Table \ref{onedemographic}, \ref{oneinterview}, \ref{onefeedback}, \ref{twofeedback}, \ref{threeobservational}, and \ref{threefeedback} list all questions, per interview round. 
Table \ref{table:results}, \ref{table:results2}, \ref{table:results3} and \ref{table:results4} summarize all answers, per question/interviewee. 
The tables also map the questions to the original RQs. 

\begin{table}[h]
\caption{Demographic Interview Questions}
\label{onedemographic}
\resizebox{\textwidth}{!}{%
\begin{tabular}{|l|l|}
\hline

Q1.1.1 & 
Please state your total industrial experience. \\\hline
Q1.1.2 &   Please rate your familiarity with the autopilot code base. Please rank out of five.      \\\hline
Q1.1.3 &
What is your current position at Micropilot?  \\\hline
Q1.1.4 & 
What is your highest level of education? \\\hline
Q1.1.5 & 
Are you familiar with state machines, in general?\\\hline

\hline
\end{tabular}%
}
\end{table}

\begin{table}[h]
\caption{Round 1 - Issue Specific Interview Questions}
\label{oneinterview}
\resizebox{\textwidth}{!}{%
\begin{tabular}{|l|l|}
\hline

Q1.2.1 & \begin{tabular}{@{}l@{}}
Explain the differences that you notice between the buggy and correct state\\ machines (e.g., number of states, sequence of functions, fields changes, etc.)\end{tabular} \\\hline
Q1.2.2 & \begin{tabular}{@{}l@{}}
After comparing both state machines, are you able to identify the field(s) that\\ caused the issue or showed the effect of the issue? \end{tabular} \\\hline
Q1.2.3 & \begin{tabular}{@{}l@{}}
Are you able to identify which function caused the issue? \end{tabular} \\\hline

\hline
\end{tabular}%
}
\end{table}

\begin{table}[h]
\caption{Round 1 - General Feedback Questions}
\label{onefeedback}
\resizebox{\textwidth}{!}{%
\begin{tabular}{|l|l|}
\hline

Q1.3.1 & \begin{tabular}{@{}l@{}}
After looking at the given list of functions and fields selected for a given issue, how\\  difficult do you think it would be if you had to select those inputs for a given\\ issue?
\end{tabular} \\\hline

Q1.3.2 & \begin{tabular}{@{}l@{}}
With respect to the selected fields and functions, are the state machines generated\\ by the tool correct? Are they detailed enough? If not, what important\\ information is missing?
\end{tabular} \\\hline

Q1.3.3 & \begin{tabular}{@{}l@{}}
For the given issues, are the set of selected fields and functions adequate? If you\\  were selecting the fields by yourself, would you have made any changes to the list\\  of selected fields and functions?
\end{tabular} \\\hline

Q1.3.4 & \begin{tabular}{@{}l@{}}
What is your current practice/procedure of debugging, when an issue is\\ reported and assigned to you? Which tool support you have?
\end{tabular} \\[10pt]\hline

Q1.3.5 & \begin{tabular}{@{}l@{}}
Assume you select the best set of fields and functions for a certain issue and use\\ our tool to generate the corresponding state machines. Do you think our tool\\
would bring any advantage to your current set of debugging/monitoring tools, at\\ Micropilot Inc.? If so what advantages and if not why?
\end{tabular} \\\hline

Q1.3.6 & \begin{tabular}{@{}l@{}}
How easy it is to embed this tool into your current frameworks and infrastructure?\\ Any challenges that you see in making this a part of your tool set?\\
\end{tabular} \\[10pt]\hline

Q1.3.7 & \begin{tabular}{@{}l@{}}
Please provide your feedback about the feasibility, effectiveness, and usability of\\ this tool and idea.
\end{tabular} \\\hline

\hline
\end{tabular}%
}
\end{table}

\begin{table}[]
\caption{Round 2 - General Feedback Interview Questions}
\label{twofeedback}
\resizebox{\textwidth}{!}{%
\begin{tabular}{|l|l|}
\hline

Q2.1 & \begin{tabular}{@{}l@{}}
How difficult was it to select the functions / fields, on a scale of one to ten, for a \\given issue?
\end{tabular} \\\hline

Q2.2 & \begin{tabular}{@{}l@{}}
After selecting the list of fields and functions, on a scale of one to ten, how \\
difficult was it to come up with the valid constraints with respect to the  scenario?
\end{tabular} \\\hline

Q2.3 & \begin{tabular}{@{}l@{}}
How many times, on average, did you have to repeat the process (reselect the fields,\\
functions select constraints, and regenerate state machine) to  abstract  the program\\ 
behavior  according  to  your  need  and  expectation?
\end{tabular} \\\hline

Q2.4 & \begin{tabular}{@{}l@{}}
How much time did it take on average, to generate the final model?
\end{tabular} \\\hline

\hline
\end{tabular}%
}
\end{table}

\begin{table}
\caption{Round 3 - Observational Questions}
\label{threeobservational}
\resizebox{\textwidth}{!}{%
\begin{tabular}{|l|l|}
\hline

Q3.1.1 & \begin{tabular}{@{}l@{}}
How many times, on average, did the subjects have to repeat the process\\
(reselect the fields, functions, and/or reselect constraints and/or \\
regenerate state machine) to create their final models?
\end{tabular} \\\hline

Q3.1.2 & \begin{tabular}{@{}l@{}}
How much time did the manual debugging take, and how much time was \\taken
by each tool run (average is reported for multiple runs)?
\end{tabular} \\\hline

Q3.1.3 & \begin{tabular}{@{}l@{}}
Which diagnostic strategies were used?
\end{tabular} \\\hline

\hline
\end{tabular}%
}
\end{table}

\begin{table}
\caption{Round 3 - Feedback Questions}
\label{threefeedback}
\resizebox{\textwidth}{!}{%
\begin{tabular}{|l|l|}
\hline

Q3.2.1 & \begin{tabular}{@{}l@{}}
How difficult was it to select the functions / fields, in general, on a scale of 1 to 10?
\end{tabular} \\\hline

Q3.2.2 & \begin{tabular}{@{}l@{}}
After selecting the list of fields and functions, on a scale of 1 to 10 how difficult was\\
it to come up with the valid constraints with respect to the scenario?
\end{tabular} \\\hline

Q3.2.3 & \begin{tabular}{@{}l@{}}
What is your current practice/procedure of debugging, when an issue is reported and\\assigned to you? Which tool support do you have?
\end{tabular} \\\hline

Q3.2.4 & \begin{tabular}{@{}l@{}}
How beneficial is using this tool compared to your current practice?
\end{tabular} \\\hline

Q3.2.5 & \begin{tabular}{@{}l@{}}
How easy is it to embed this tool into your current frameworks and infrastructure? Any\\ challenges that you see in making this a part of your tool-set? 
\end{tabular} \\\hline

Q3.2.6 & \begin{tabular}{@{}l@{}}
Please provide your feedback about the feasibility, effectiveness, and usability of this\\
tool and the idea of using your knowledge to break the issues down to fields, functions\\and constraints.
\end{tabular} \\\hline

\hline
\end{tabular}%
}
\end{table}

\begin{table}[]
\caption{Summarized responses to interview questions in Rounds 1 and 2 (Part 1). F: Feasibility, E: Effectiveness, U:Usability}
\label{table:results}
  \begin{center}
    \begin{tabular}{ | c | c | c | c |}
      \hline
      \thead{ID} & \thead{Subject 1\\(Round 1)} & \thead{Subject 2\\(Round 1)}& \thead{Subject 3\\(Round 1,2)} \\
      \hline

       \makecell{Industry\\Experience (yrs)} & 2 & 0.9 & 2.5 \\\hline

\makecell{Code Base\\Experience (/5)} & 3 & 2 & 2 \\\hline

Q1.2.1 & all diff. spotted & all diff. spotted & all diff. spotted  \\\hline

Q1.2.2 & 5/5 correct fields & 4/5 correct fields & 4/5 correct fields \\\hline

Q1.2.3 & 5/5 correct func. & 4/5 correct func. & 4/5 correct func.  \\\hline

Q1.3.1 (/5) & 2 & 2 & \makecell{4 if not familiar} \\\hline

Q1.3.2 & Correct & Correct & Correct  \\\hline

Q1.3.3 & Adequate & Adequate & Would add more \\\hline

Q1.3.4 & \makecell{Reproduce, debug \\with Visual Studio} & \makecell{Reproduce, simulate, \\debug with Visual\\-Studio} & \makecell{ Reproduce, debug \\with Visual Studio}  \\\hline

Q1.3.5 & Visual component & \makecell{Breaking debug \\activity to functions \\to get starting point} & \makecell{Big picture,  \\suggests the \\ functions we \\ should review}  \\\hline

Q1.3.6 & \makecell{Seems usable, \\no major problems} & \makecell{Makes debugging \\faster, worthy \\to get used to}  & \makecell{Selecting fields is \\easier, coming up \\with constraints is \\difficult, not hard \\to get used to}   \\\hline

Q1.3.7 & F=9, E=8, U=8 & F=7, E=7, U=7 & F=9, E=9, U=9 \\ \hline

Q2.1 (out of 10) & N/A & N/A & 2 \\ \hline

Q2.2 (out of 10) & N/A & N/A & 2 \\ \hline

Q2.3 & N/A & N/A & One \\ \hline

Q2.4 & N/A & N/A & $<$ 10 min \\ \hline

    \end{tabular}
  \end{center}
\end{table}


\begin{table}[]
\caption{Summarized responses to interview questions in Rounds 1 and 2 (Part 2). F: Feasibility, E: Effectiveness, U:Usability}
\label{table:results2}
  \begin{center}
    \begin{tabular}{ | c | c | c | c |}
      \hline
       
\thead{ID} & \thead{Subject 4\\(Round 1,2)}  & \thead{Subject 5\\(Round 1)} & \thead{Mapped to\\ (RQ)} \\
& & &\\\hline
\makecell{Industry\\Experience (yrs)} & 0.9 & 1.5 & N/A \\\hline
\makecell{Code Base\\Experience (/5)} & 2 & 2 & N/A\\\hline
Q1.2.1 & all diff. spotted & all diff. spotted & RQ1 \\\hline
Q1.2.2 & 5/5 correct fields & 5/5 correct fields & RQ1 \\\hline
Q1.2.3 & 5/5 correct func. & 5/5 correct func. & RQ1 \\\hline
Q1.3.1 (/5) & 2 & 2 & RQ2 \\\hline
Q1.3.2 & \makecell{Missing details \\(branching inside \\methods, conditions \\etc)} & Correct & RQ1 \\\hline
Q1.3.3 & Would add more & Would add more & RQ2 \\\hline
Q1.3.4 & \makecell{Reproduce, debug \\with Visual Studio} & \makecell{Reproduce, debug \\with Visual Studio} & RQ3 \\\hline
Q1.3.5 & \makecell{Visual component, \\suggests the functions \\we should review} & \makecell{Somewhat advanta-\\-geous, doesn't state \\the line of issue}  & RQ3 \\\hline
Q1.3.6 & \makecell{Easy to learn \\and review} & No challenges & RQ4  \\\hline
Q1.3.7 & F=9, E=6, U=9 & F=7, E=9, U=8 & \makecell{RQ2, RQ3, \\RQ4}\\ \hline
Q2.1 (out of 10) & 2 & N/A & RQ2 \\ \hline
Q2.2 (out of 10) & 2 & N/A & RQ2 \\ \hline
Q2.3 & One & N/A & RQ2 \\ \hline
Q2.4 & $<$ 10 min & N/A & RQ2 \\ \hline  
       
    \end{tabular}
  \end{center}
\end{table}


\begin{table}[]
\caption{Summarized responses to interview questions in Round 3 (Part 1). F: Feasibility, E: Effectiveness, U:Usability}
\label{table:results3}
  \begin{center}
    \begin{tabular}{ | c | c | c | c |}
      \hline
       
\thead{ID} & 
\thead{Subject 6\\(Round 3)} & 
\thead{Subject 7\\(Round 3)}  \\\hline

\makecell{Industry\\Experience (yrs)} & 2 & 5  \\\hline
\makecell{Code Base\\Experience (/5)} & 0 & 2  \\\hline
Q3.1.1 & 1 & 1  \\ \hline
Q3.1.2 & \makecell{Manual: 25 min \\ Tool: 12 min} & \makecell{Manual: 12 min \\ Tool: 10 min}  \\ \hline
Q3.1.3 & \makecell{Code Comprehension, Forward\\Reasoning} & Code Comprehension \\ \hline
Q3.2.1 (out of 10) & 1 & 3   \\\hline
Q3.2.2 (out of 10) & 1 & 3    \\\hline
Q3.2.3 & \makecell{Debug. Approach: Code Comprehension,\\  Forward  Reasoning\\Toolset: Visual Studio} & \makecell{Debug. Approach: \\Code Comprehension, \\Forward Reasoning, \\Backward Reasoning\\Toolset: Visual \\Studio} 
\\\hline
Q3.2.4 & \makecell{Beneficial in identifying location of \\ bugs in complex scenarios}  & \makecell{Beneficial as easily \\traces and points to \\the location of bug} \\\hline
Q3.2.5 & No challenges & No challenges  \\\hline
Q3.2.6 & F=10, E=9, U=8 & F=9, E=8, U=9 \\ \hline

    \end{tabular}
  \end{center}
\end{table}


\begin{table}[]
\caption{Summarized responses to interview questions in Round 3 (Part 2). F: Feasibility, E: Effectiveness, U:Usability}
\label{table:results4}
  \begin{center}
    \begin{tabular}{ | c | c | c | c |}
      \hline

\thead{ID} & 
\thead{Subject 8\\(Round 3)} & 
\thead{Mapped To (RQ)\\(Round 3)}  \\\hline

\makecell{Industry\\Experience (yrs)} &  13 & N/A\\\hline
\makecell{Code Base\\Experience (/5)} & 5 & N/A\\\hline
Q3.1.1 & 1 & RQ2\\ \hline
Q3.1.2 & \makecell{Manual: 8 min \\ Tool: 11 min} & RQ2\\ \hline

Q3.1.3 & Backward Reasoning & RQ3 \\\hline
Q3.2.1 (out of 10) & 3 & RQ2 \\\hline
Q3.2.2 (out of 10) & 3 & RQ2 \\\hline
Q3.2.3 & \makecell{Debug. Approach: Offline\\ Analysis, Input Manipulation\\Toolset: Visual Studio, Valgrind}  & RQ3\\\hline
Q3.2.4 & \makecell{Beneficial in analyzing system \\behavior and verifying effects of\\code changes} & RQ3 \\\hline
Q3.2.5 & Tool Maintenance & RQ4 \\\hline
Q3.2.6 & F=N/A, E=8, U=8 & RQ2, RQ3, RQ4 \\ \hline

    \end{tabular}
  \end{center}
\end{table}


\end{document}